\documentclass[11pt, oneside]{article}
\usepackage{geometry}
\usepackage{graphicx}		
\usepackage{lipsum}
\usepackage{comment}
\usepackage{amsmath}
\usepackage{url}
\usepackage{multirow}

\newcommand{\TitledBox}[2]{%
    \vspace{1em} % Adds vertical space before the box
    \noindent
    \fbox{%
        \parbox{\dimexpr\linewidth-2\fboxsep-2\fboxrule}{%
            \textbf{#1}\\[0.5em] % Title in bold, with a little space below
            #2 % Content of the box
        }%
    }%
    \vspace{1em} % Adds vertical space after the box
}

\title{\textbf{Group size effects and collective misalignment
\\ in LLM multi-agent systems}}

\author{% You can write out first names or use initials - either way is acceptable, but be consistent
    Ariel~Flint$^{1}$,
	Luca~Maria~Aiello$^{2,3}$,
    Romualdo~Pastor-Satorras$^{4}$,
	Andrea~Baronchelli$^{1\ast}$\and \and \\
	\small$^{1}$Department of Mathematics, City St George’s University of London, UK. \\
	\small$^{2}$IT University of Copenhagen, Denmark.\\
    \small$^{3}$Pioneer Centre for AI, Copenhagen, Denmark. \\
    \small$^{4}$Departament de Física, Universitat Politècnica de Catalunya, Barcelona, Spain. \\
	\small$^\ast$Corresponding author. Email: a.baronchelli.work@gmail.com}

\date{}
\begin{document}
\maketitle

\begin{abstract}
\noindent Multi-agent systems of large language models (LLMs) are rapidly expanding across domains, introducing dynamics not captured by single-agent evaluations. Yet, existing work has mostly contrasted the behavior of a single agent with that of a collective of fixed size, leaving open a central question: how does group size shape dynamics? Here, we move beyond this dichotomy and systematically explore outcomes across the full range of group sizes. We focus on multi-agent misalignment, building on recent evidence that interacting LLMs playing a simple coordination game can generate collective biases absent in individual models. First, we show that collective bias is a deeper phenomenon than previously assessed: interaction can amplify individual biases, introduce new ones, or override model-level preferences. Second, we demonstrate that group size affects the dynamics in a non-linear way, revealing model-dependent dynamical regimes. Finally, we develop a mean-field analytical approach and show that, above a critical population size, simulations converge to deterministic predictions that expose the basins of attraction of competing equilibria. These findings establish group size as a key driver of multi-agent dynamics and highlight the need to consider population-level effects when deploying LLM-based systems at scale.
\end{abstract}
%\vspace{0.5cm}

\section*{Introduction}
Multi-agent LLM systems have recently started to be deployed and are already responsible for tasks in domains such as finance \cite{ferreira2021artificial,sun2023reinforcement}, defense \cite{black2024strategic}, energy \cite{camacho2024leveraging}, social media \cite{schroeder_2025}, and personal assistance \cite{li2024personal}. These applications are expected to expand rapidly, as reflected in growing market interest, with valuations projected to increase from USD 5.1 billion in 2024 to USD 47.1 billion by 2030 and a surge in startups focused on interacting LLM agents \cite{springs2024multiagent,grandview2024agentic}. Scalability, autonomy, and interaction, which are at the root of the success of LLM-based multi-agent systems, also give rise to new risks. Building on evidence that multi-agent safety is not guaranteed by single-agent safety \cite{anwar2024foundational}, the International AI Safety Report 2025, authored by 100 experts including representatives from 33 countries and intergovernmental organizations, has warned of the unprecedented complexity of multi-agent LLM systems \cite{AI_Safety_Report_2025}, highlighting risks such as systemic failures, misaligned collective behaviors, and uncontrollable strategic dynamics \cite{CAIF_1}. As multi-agent LLM systems become a dominant paradigm in AI infrastructure, understanding their dynamics, including both capabilities and failure modes, is widely recognized as a central scientific and governance priority \cite{brinkmann2023machine,tsvetkova2024new}.

Recent research has shown that local interactions among agents can yield collective behaviors and performance gains that are not observed when models operate in isolation \cite{guo2024large,mou2024individual}. For example, at the smallest scale, dyads (pairs of agents) engaging in debate lead to substantial improvements in factual accuracy and truthfulness \cite{du2023improving}, and may also enhance coherence and depth of LLM answers~\cite{khan2024debating}. Triads of LLMs have been shown to produce more utilitarian collective moral judgments than individual models \cite{keshmirian2025many}, and may optimize collaboration in complex tasks~\cite{zhang2023exploring}. Small groups of agents can accomplish goals more effectively through self-organized division of labor \cite{du2023review}. Groups of tens of agents are able to develop social conventions  \cite{ashery2025emergent}, and to autonomously coordinate social behaviors that mirror aspects of human interaction in immersive environments \cite{park2023generative}. Populations on the order of hundreds or thousands have been used to investigate mechanisms of cultural evolution \cite{perez2024cultural,de2024ai,park2024generative}, while even larger populations, scaling from tens to hundreds of thousands, have been studied as in vitro societies \cite{piao2025agentsociety}. Frameworks have also been proposed to model the dynamics of millions \cite{yang2024oasis}, or even billions \cite{guan2025modeling}, of interacting LLMs.

While convincingly supporting the idea that \textit{more is different} for populations of LLMs \cite{anderson1972more}, the vast majority of studies in this rapidly growing body of work are characterized by a recurring methodological structure: the behavior of a population is compared with that of a single agent, revealing differences between the two conditions. However, this approach leaves a fundamental question unanswered: what happens across varying population sizes? In other words, what does \textit{more} quantitatively mean? This question is crucial for applications involving scalable deployments of LLM agents, where it is critical to understand whether there exists a threshold beyond which increasing the population size no longer affects system behavior, whether in terms of performance, coordination, or emergent properties.

Here, we investigate the role of group size on LLM collective dynamics by focusing on bias, a critical risk associated with language models that has been extensively studied in isolated settings and carries far-reaching implications for safety and alignment ~\cite{roselli2019managing,ferrara2023should,hu2023generative,carichon2025comingcrisismultiagentmisalignment}. We build on recent empirical observations showing that groups of LLMs coordinating on shared linguistic conventions can exhibit collective misalignment~\cite{ashery2025emergent}, introducing bias even when individual agents are unbiased. In this context, individual bias refers to an initial statistical preference for one option over an equivalent alternative during coordination (for example, agents systematically preferring one name over another in a naming process). Collective bias, by contrast, denotes an additional imbalance that emerges from interactions among multiple LLMs, increasing the likelihood that certain coordination outcomes are adopted over others that are, in principle, equally viable. 

The contribution of this paper is threefold. We clarify and extend the phenomenon of collective bias as a form of collective misalignment, showing how coordination can amplify, create, or overturn individual biases. We map how the strength and form of collective bias depend on group size, identifying non-linear effects and qualitative shifts in the dynamics. We provide an analytical framework that clarifies why simulations converge beyond a given population size and how basins of attraction structure the competition between equilibria. These findings are robust across different LLM models.

\section*{Modeling Framework and Parameterization}

\subsection*{Modeling Framework} 
We consider a population of $N$ LLM agents engaged in a naming game~\cite{steels1995self,baronchelli2006sharp}, a standard framework for studying the emergence of conventions that has been extensively investigated through theoretical modelling~\cite{castellano2009statistical,baronchelli2018emergence} and laboratory experiments with humans~\cite{centola2015spontaneous,centola2018experimental}.
At each time step, two agents are randomly selected to interact, producing and exchanging a convention, or word, chosen from a finite pool of size $W$, with the goal of maximizing their respective game scores (see Methods and SI for prompting and meta-prompting details). The outcome of an interaction yields identical payoffs to both agents. If both agents select the same convention, their scores increase by the same amount; otherwise, they decrease by the same amount. The reward for successful coordination exceeds the penalty for failure. This scoring rule incentivizes local coordination between pairs of agents, without any explicit drive toward global consensus. Each agent maintains a finite memory state $M$ of capacity $H$, which stores information about its most recent interactions: its own and its partner’s convention choices, whether coordination was achieved, and the accumulated score over the last $H$ encounters. At the beginning, memories are empty, and the first choice is drawn from the available pool according to the agent’s individual bias. Recent analyses of this protocol in LLM populations reveal that such local interactions can spontaneously lead to global consensus on a specific word and, crucially, generate collective bias even when individual agents exhibit none \cite{ashery2025emergent}.

\subsection*{Implementation}
The modeling framework described above can be directly implemented by running large language models (LLMs) as agents, where the memory state of each agent is translated into a text prompt that the LLM uses to generate a new convention~\cite{ashery2025emergent}. To study the coordination dynamics of large populations of LLM agents, however, we introduce a complementary stochastic model that enables simulations with population sizes far beyond what is computationally feasible in direct LLM experiments, while drastically reducing the required computational resources.

In this stochastic model, an agent’s behavior is represented by a \emph{probabilistic policy} that specifies the likelihood of producing each of the $W$ possible words based on the agent’s current memory state. We derive this policy empirically by probing the LLM: given a textual prompt encoding a particular memory state, the LLM outputs a vector of logits indicating the likelihood of all possible next tokens. We extract and normalize the logits corresponding to our restricted set of $W$ words to form a probability distribution, which defines the agent’s policy for that memory state (see Methods for full details on  model definition and policy extraction procedure).

To enable efficient large-scale simulations, we precompute and store these probabilistic policies for all possible memory states. During simulation runs, agents sample conventions according to their stored policies, thereby capturing the stochastic nature of LLM-generated interactions without invoking the underlying model at every step. This approach is validated by extensive experiments showing close agreement between direct LLM-based generations and policy-driven simulations (see SI for quantitative comparisons, particularly Figs.~\ref{fig: original vs new method} and~\ref{fig: matrix test}).

\subsection*{Parametrization}
In this paper, we restrict our analysis to the case of $W=2$ competing conventions. Since we use multi-agent bias as the main measurement to explore size effects, we select pairs of words previously identified as particularly sensitive to bias (e.g., \{\textit{man, woman}\} or \{\textit{straight, gay}\})~\cite{nangia_crows-pairs:_2020}. %
To test robustness across architectures, we feed the model with probability policies extracted from different instruction-tuned LLMs:
Qwen QwQ-32B (denoted as Qwen), Microsoft Phi-4 (Phi), OpenAI GPT-4o (GPT), and Meta Llama 3.1 Instruct (Llama). 

We consider only homogeneous populations, meaning that the probabilistic policies of all agents are extracted from the same underlying LLM model. Finally, thorough the paper, time is measured in population rounds (Monte-Carlo steps), each consisting of $N$ pairwise interactions. Simulations run for up to 1000 population rounds or until all agents converge on the same convention. Theoretical predictions for simulations are based on the minimal naming game model \cite{baronchelli2006sharp}, in which agents may additionally be individually biased towards one convention (see SI).

\section*{Results}

\subsection*{Mapping collective misalignment}

\begin{figure}[!t]
    \centering
    \includegraphics[width=0.7\linewidth]{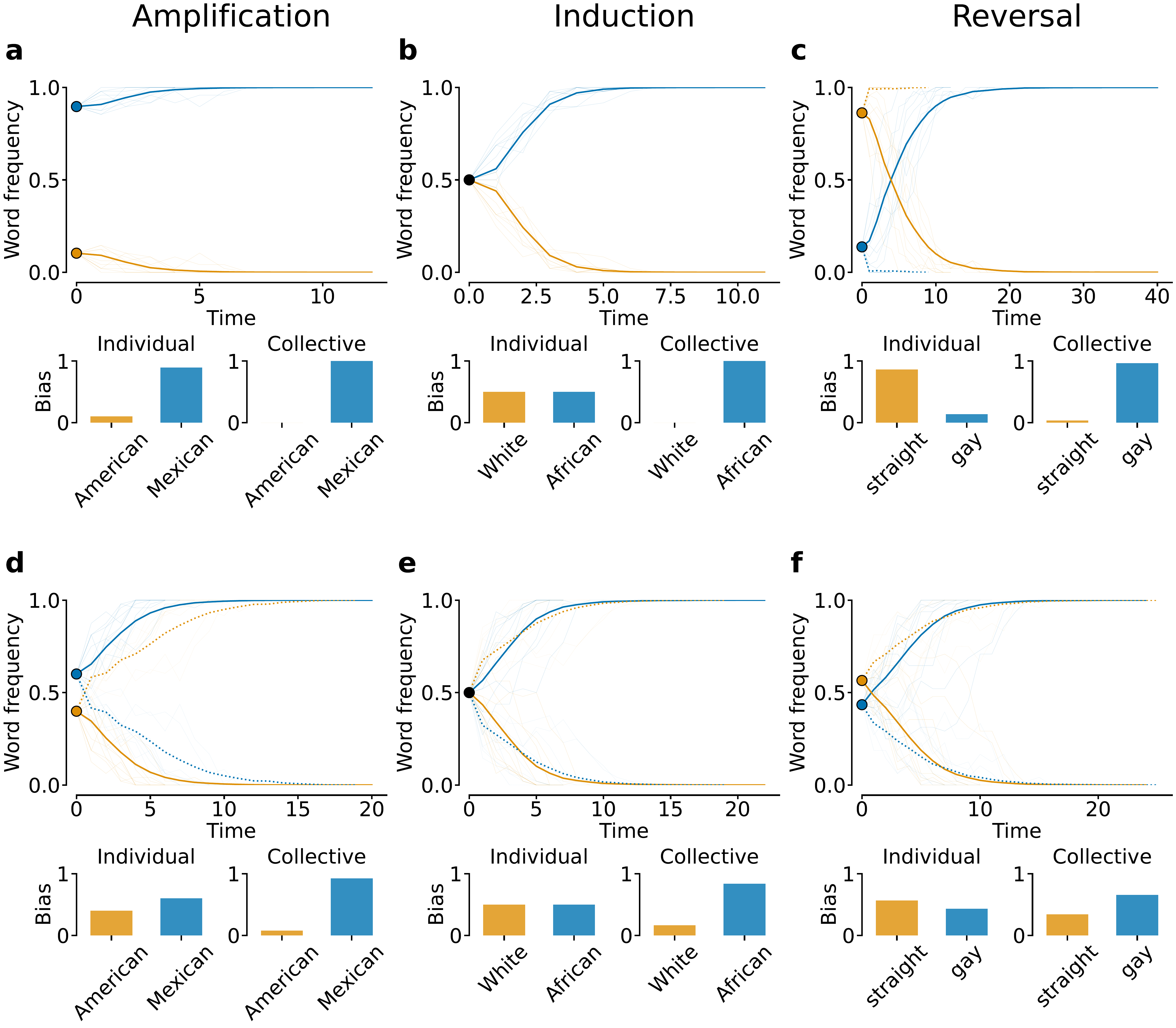}
    \caption{\textbf{Interaction can amplify, induce, or override individual bias.} From left to right columns, the word pairs that populations coordinate on are: \{\textit{American, Mexican}\}, \{\textit{White, African}\}, and \{\textit{straight, gay}\}. These cases illustrate bias amplification, induction from neutrality, and bias reversal, respectively. Top row (a–c): Llama populations; bottom row (d–f): GPT populations. All simulations use a population size of $N=24$. The upper plots in each panel display representative trajectories of word competition in 1000 simulation runs, where blue and orange lines indicate the frequency of unique convention choices over time (based on the previous $N$ interactions). 
    Colored circles at $t=0$ denote the initial individual bias, with black indicating neutrality. 
    Where possible, at least ten trajectories are shown for each consensus outcome (strong or weak convention); when fewer runs converged, all available trajectories are displayed. 
    Solid and dotted lines show the mean dynamics of runs that converged on the strong and weak conventions, respectively. 
    In all cases, the strong word is represented by blue trajectories (from left to right: \textit{Mexican}, \textit{African}, \textit{gay}). 
    The lower bar plots summarize individual and collective bias: individual bias reflects agents’ pre-interaction preferences, and collective bias shows the fraction of runs that reached consensus on each convention.
}
    \label{fig: intro}
\end{figure}

We begin by examining how individual bias shapes collective outcomes by performing simulations of our model for small system sizes. Fig.~\ref{fig: intro} shows that individual biases are not necessarily reflected by collective outcomes. Instead, interaction gives rise to three distinct forms of misalignment between individual and collective bias. First, individual preferences may be amplified (Fig.~\ref{fig: intro}a,d): collective dynamics increase the likelihood of consensus on the word agents initially favor. Second, neutral individuals can give rise to collective bias, consistent with previous findings~\cite{ashery2025emergent} (Fig.~\ref{fig: intro}b,e). Finally, collective dynamics can override individual preferences, producing consensus on the word agents initially disfavor (Fig.~\ref{fig: intro}c,f).

\begin{figure}[!t]
    \centering
    \includegraphics[width=0.5\linewidth]{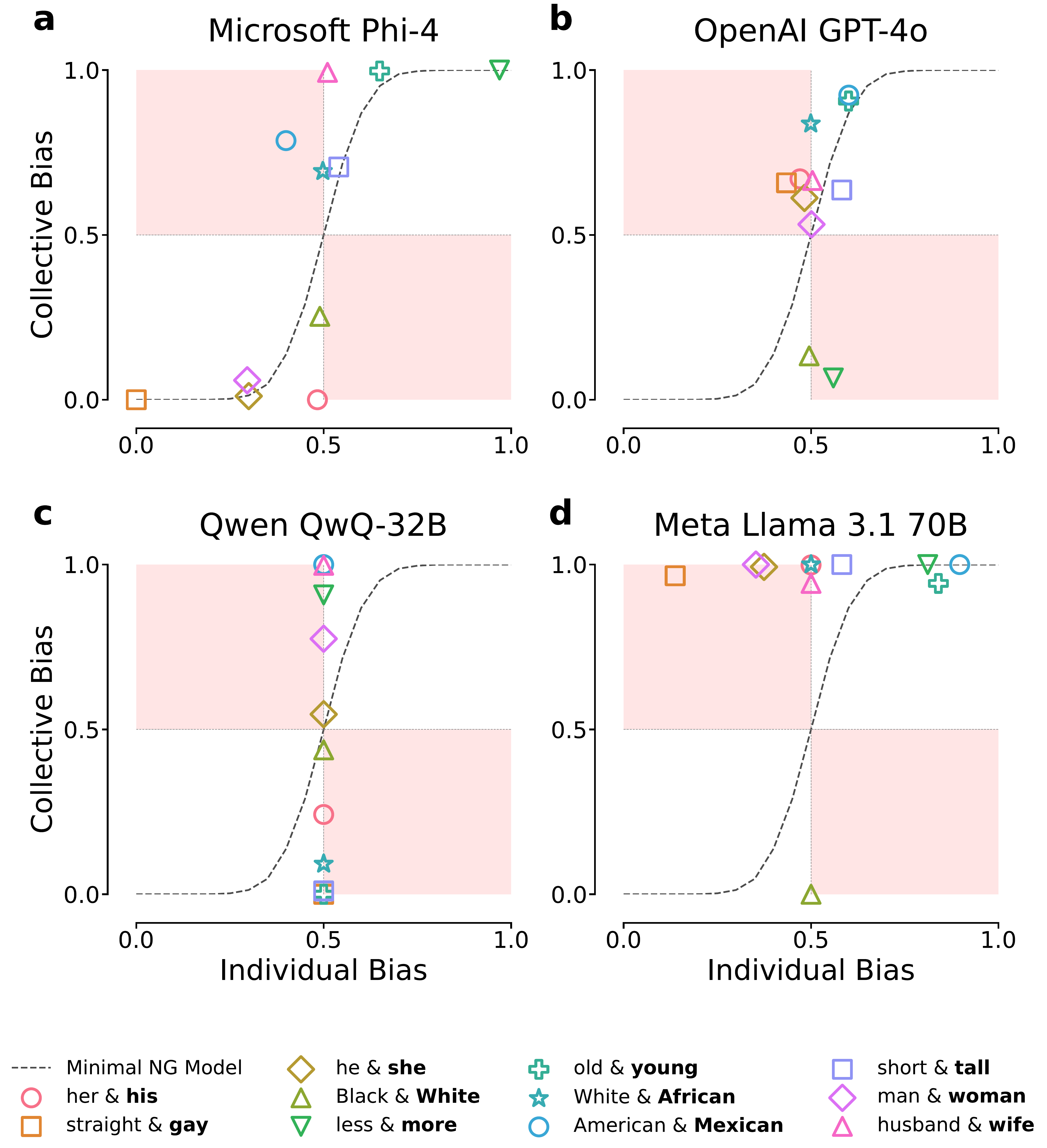}
    \caption{\textbf{Collective bias depends on both model and convention.} Panels show individual and collective bias for four LLMs (clockwise from top left: Phi, GPT, Qwen, and Llama) in populations of size $N=24$, for the word pairs indicated in the legend. Individual bias corresponds to an agent’s selection probability at the start of the game with empty memory, while collective bias denotes the proportion of 1000 runs that reached consensus on the option in bold. Error bars (SEM) are smaller than the marker size. The dashed line shows the theoretical prediction from the minimal Naming Game with binary options and individual bias. Points along the gray dotted line at $x = 0.5$ indicate symmetry breaking, where neutral individual preferences produce biased collective outcomes. Points within the shaded pink regions correspond to bias reversal, where collective interactions overturn the individual preference.}
    \label{fig: collective vs individual bias pairs}
\end{figure}

To systematically assess the robustness of these misalignment patterns, we extend the analysis across eleven word pairs and four homogeneous LLM populations. Fig.~\ref{fig: collective vs individual bias pairs} contrasts individual and collective biases for four LLM models in populations of size $N=24$, revealing strong variations across LLM models and word pairs. The dashed black line indicates theoretical predictions from the minimal naming game model, which assumes agents with fixed, memory-invariant bias. LLM agents, by comparison, adjust according to the memory context.

At the individual level, where single agents have empty memory states prior to interaction, Phi and Llama display considerable variability in individual bias, whereas GPT and Qwen are closer to neutrality. At the collective level, however, coordination outcomes differ markedly across models. Qwen populations exhibit a wide spectrum of collective bias strengths, GPT populations show only mild collective effects, while Llama populations are highly polarized and converge almost deterministically on a single word for each pair. Phi populations most closely align with the minimal naming game model predictions, though they still diverge strongly in five of the eleven word pairs. 

Importantly, both the magnitude and the direction of collective bias can differ across models. For example, for the pair \{\textit{her, his}\}, populations of Qwen and Phi converge on \textit{her}, while GPT and Llama converge on \textit{his}, despite nearly identical individual-level tendencies.

\subsection*{The role of population size} 
\begin{figure}[!t]
    \centering
    \includegraphics[width=0.5\linewidth]{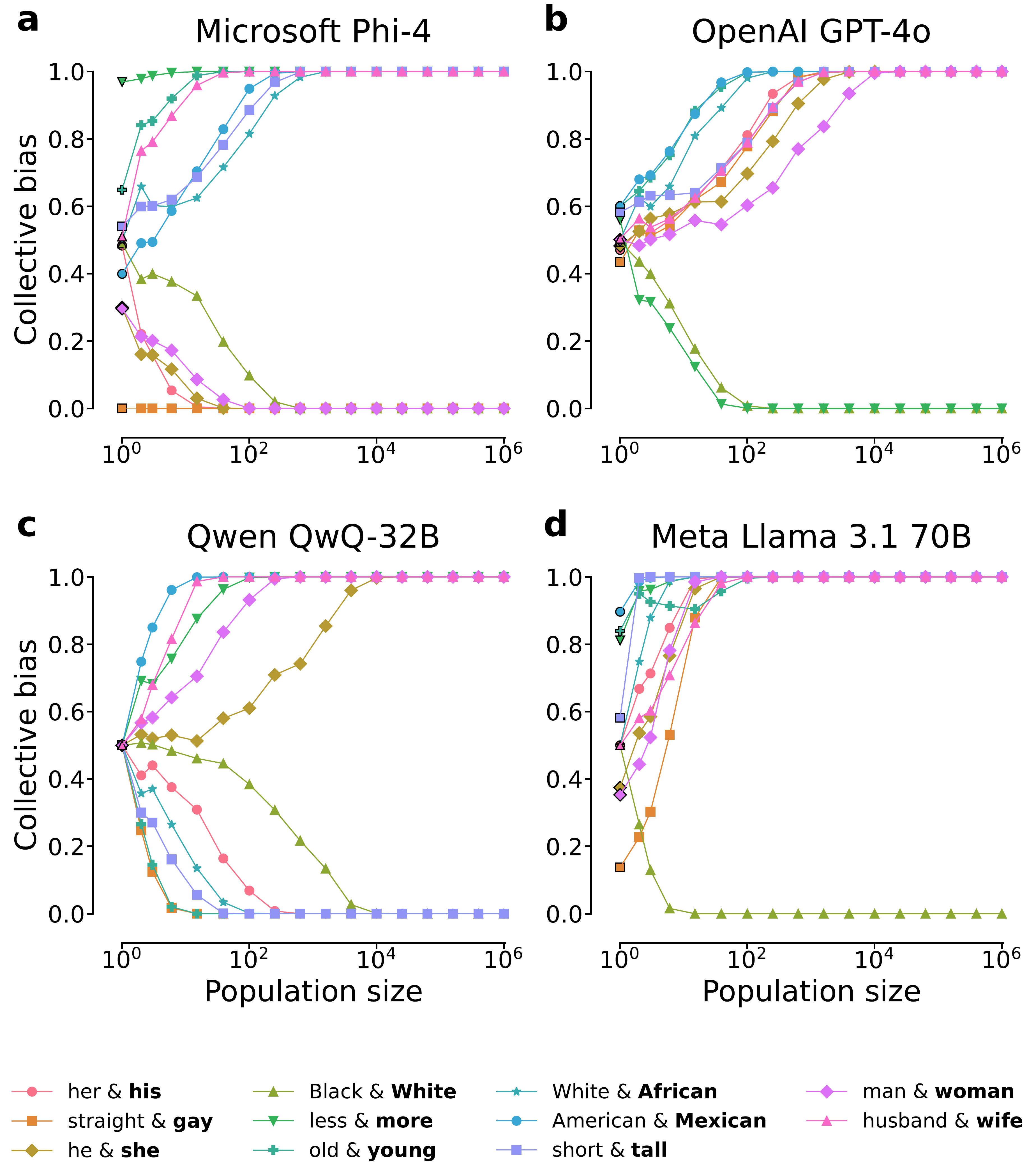}
    \caption{\textbf{Population size effects on collective bias.} 
    For all convention pairs tested, collective preference for a given convention increases with population size until convergence becomes deterministic. The individual bias is reported at $N=1$. Error bars (SEM) are smaller than the marker size. Across population sizes, all runs reached consensus except in three cases: in Llama populations, the convergence rate decreases beyond a certain $N$ for \textit{\{old, young\}} and \textit{\{less, more\}}, and in small Qwen populations, a few runs failed to converge for the word pair \textit{\{husband, wife\}}. See SI Figs.~\ref{fig: non-consensus_fraction_410}-\ref{fig: non-consensus_dynamics_1161} for word competition dynamics and precise convergence fractions for these cases.}
    \label{fig: population size}
\end{figure}

Having established that LLM populations can exhibit diverse collective behavior, we next examine the role of population size in shaping these outcomes. Fig.~\ref{fig: population size} shows that the probability of converging on a specific word grows systematically with $N$ across all models and word pairs. In other words, larger populations reduce outcome uncertainty. We label the preferred word as the ``strong'' word, as opposed to its ``weak'' alternative. Once $N$ exceeds a threshold, the collective outcome becomes fully deterministic: if the population can reach consensus, then it will always converge on the strong word. 

The threshold varies considerably across model and word pair combinations. In some cases, determinism arises for populations as small as $N=2$ (Llama, \{\textit{short}, \textit{tall}\}), while in others, finite-size effects persist up to $N\sim10^4$ (Qwen, \{\textit{Black}, \textit{White}\}). The rate at which collective bias grows with $N$ likewise depends on both the LLM and the word pair. Figure~\ref{fig: population size}c shows that although Qwen exhibits uniform strategies at the individual level ($N=1$) for all word pairs, the magnitude of collective bias at intermediate $N$ values varies. Population size can also modulate the form of multi-agent misalignment: for the word pair \{\textit{straight}, \textit{gay}\}, Llama shows an individual preference for \textit{straight}, but reversal toward \textit{gay} occurs only for $N\geq6$. 
Finally, we note that populations always converge, except large Llama populations for the word pairs \{\textit{old}, \textit{young}\} and \{\textit{less}, \textit{more}\}, and a minor fraction of simulation runs for small Qwen populations coordinating on the word pair \{husband, wife\}. We document these in the SI and defer systematic analysis to future  (see SI Figs.~\ref{fig: non-consensus_fraction_410}-\ref{fig: non-consensus_dynamics_1161}).

\subsection*{Convergence time}
\begin{figure}[!t]
\centering
\includegraphics[width=0.8\linewidth]{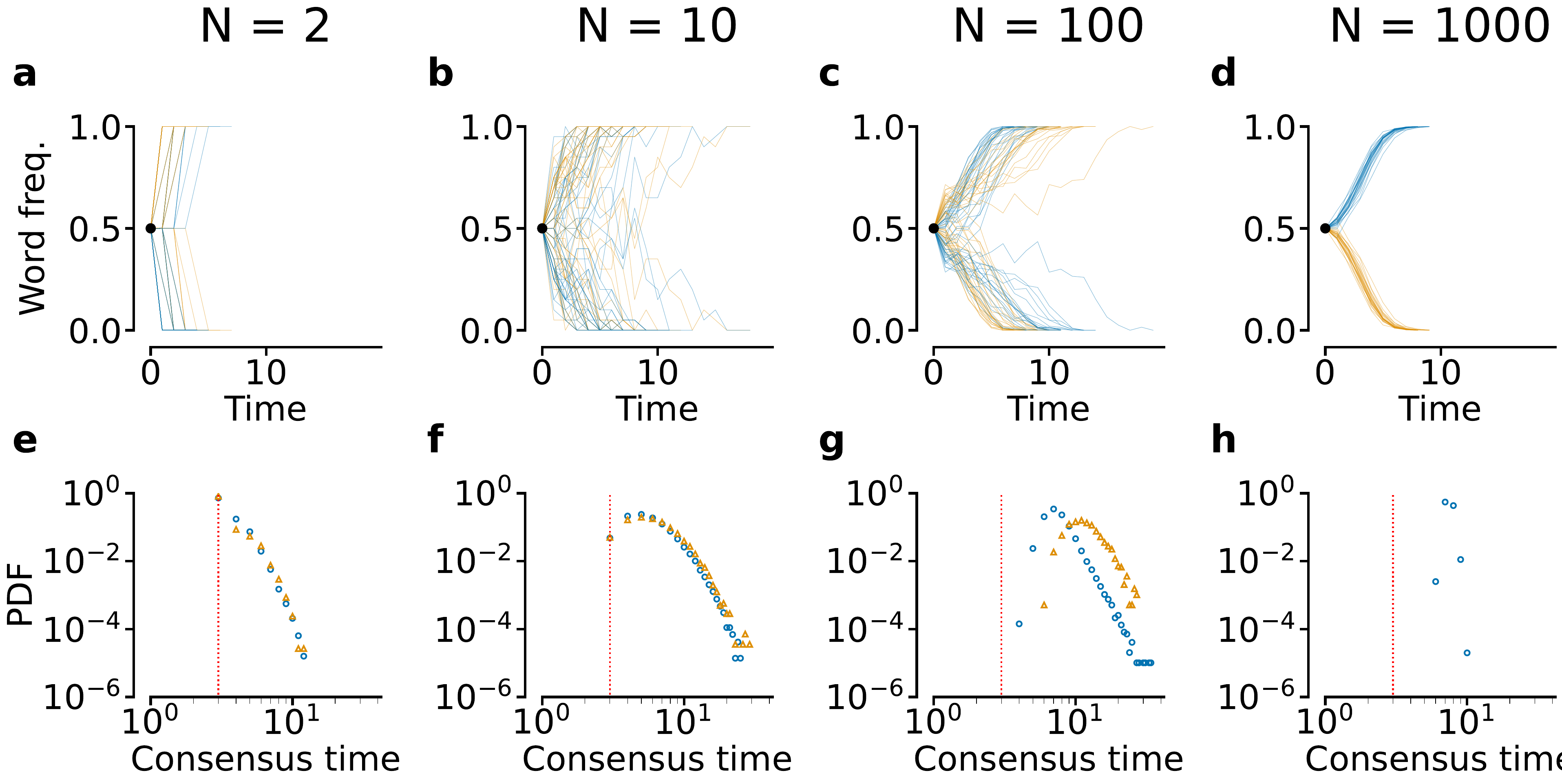}
\caption{\textbf{Coordination dynamics become increasingly deterministic for larger populations.} Each column shows the temporal dynamics of word competition (top row), and the probability density function of the consensus time (bottom row) for trajectories that reached consensus. Specifically, each column shows the results at different population sizes, for GPT populations coordinating on the word pair \{\textit{White}, \textit{African}\}. The PDFs in the bottom row are constructed from 100,000 observations in total, where blue circles and orange triangles correspond to trajectories that converged on the strong (\textit{African}) and weak (\textit{White}) word, respectively. The dotted red line indicates the shortest possible consensus time allowed by the convergence criterion, $t=3$. The top row shows the usage fraction at each population round of the strong (blue lines) and weak (orange lines) words in up to 25 trajectories for each consensus state. The collective bias toward the word \textit{African} grows with population size as (from left to right columns, up to 3 s.f.): 0.626, 0.720, 0.981, 1.00. }\label{fig:4}
\end{figure}

The dependence of collective bias on population size can be explained by qualitative shifts in the underlying consensus dynamics. In small populations, early random fluctuations dominate the dynamics. A word that gains even a small initial advantage is quickly incorporated into the memory inventories of most agents, driving rapid alignment across the system. As shown in Fig.~\ref{fig:4}, trajectories collapse to consensus (on either word) within only a few population rounds. Because the likelihood of a word becoming an early favorite depends on the agents’ initial individual biases, the collective outcome ultimately reflects these underlying preferences.

In intermediate size populations, initial fluctuations are rarely sufficiently strong to produce the rapid alignment observed at small $N$. Consensus arises from the interplay of stochastic fluctuations, which may affect the prevalence of one word even beyond the initial phase, and coordination dynamics, which tends to favor the strong word, steering trajectories toward its absorbing state. This coexistence of rapid collapses and prolonged mixed states produces a unimodal distribution of consensus times with a heavy right tail (Fig.~\ref{fig:4}).

As populations grow even larger, the consensus-time distributions separate: the average consensus time for trajectories converging on the weak word becomes larger than the consensus time to the strong word (see Figs.~\ref{fig: mode_time_llama}-\ref{fig: mode_time_phi}). This widening gap in characteristic convergence time suggests that it becomes increasingly difficult for the population to converge on the weak word. Beyond the threshold size $N_c$, the consensus state becomes deterministic: for a given word pair and model, all simulations converge on the strong word and the characteristic consensus time stabilizes (see Figs.~\ref{fig: bias_time_llama}-\ref{fig: bias_time_phi}). The consensus-time distribution for trajectories that converge on the strong word (the only viable consensus state) narrows, until $N$ is so large that nearly all trajectories converge at this characteristic convergence time.

The coordination dynamics depend on the model and the word pair. In practice, this means that while the qualitative shift from fluctuation-driven to coordination-driven consensus is robust, the transformation stages of the consensus time PDF can differ in duration and form across experimental conditions (see SI Figs.~\ref{fig: PDF_191}-\ref{fig: PDF_410}).

\subsection*{Mean-field theory} 

To understand the deterministic outcomes observed in large populations, we develop a mean-field theory that captures the system dynamics in the limit $N \to \infty$. 
Mapping the dynamics of the experimental framework to a reaction-diffusion process (see SI), we define the system state as $\mathbf{x} = \{x_i\}$, where $x_i = N_i/N$ represents the fraction of agents in state $i$ at a given timestep. The evolution of the system is characterized by the following mean-field rate equation:
\begin{equation}
  \label{eq:25}
  \frac{d x_k} {dt} = - x_k +\sum_i \sum_j x_i x_j P_k(i, j),
\end{equation}
where $P_k(i, j)$ is the probability that an agent in memory state $i$ will transition to a new memory configuration $k$ after interacting with an agent in memory state $j$. These dynamics are characterized by fixed points given by the algebraic equation
\begin{equation}
  \label{eq:17main}
  x_k  = \sum_i \sum_j x_i x_j P_k(i, j).
\end{equation}
These fixed points take the form of homogeneous absorbing states, corresponding to full consensus on one of the two available options. That is, the memory of all agents is equal to $2H$ repetitions of the first word, or $2H$ repetitions of the second word. 

We therefore denote them as the strong and weak fixed points.
Although, in principle, other, more complex  mixed fixed points could exist, they are not generally observed in numerical solutions of the mean-field equations. 
Only for a few specific combinations of LLM models and word pairs (i.e. \{\textit{old}, \textit{young}\} and \{\textit{less}, \textit{more}\} in large Llama populations) do mixed configurations appear, corresponding to time-dependent steady states also observed in direct simulations of the agent-based model at finite $N$. The mean-field dynamics for all models are shown in Figs.~\ref{fig: mean_field_phi}-\ref{fig: mean_field_llama}.

A linear stability analysis allows to determine if the homogeneous fixed points are stable or not by looking at the sign of the largest eigenvalue of the corresponding Jacobian matrix. In Table~\ref{tab:combined_eigenvalues}, we report the largest eigenvalues computed for the weak and strong fixed points. As we can see, for most combinations of LLM model and word pair either the strong fixed point is stable and the weak one unstable, or both are stable. In this last case, the actual preference for the strong word is due to the interplay of the basins of attraction of the two fixed points.

Finally, this analysis explains the observed cases of non-consensus for the word pairs \{\textit{less}, \textit{more}\} and \{\textit{old}, \textit{young}\} Llama populations. In the first case, both homogeneous fixed points are unstable, while in the second one the strong fixed point is unstable and the weak one is marginal, with an almost zero largest eigenvalue. For the pair \{\textit{husband}, \textit{wife}\} in Qwen populations, the weak fixed point in unstable and the strong one marginal, so in this case we also observe a non-homogeneous fixed point.

\section*{Discussion}
In this work, we have shown that interactions among LLM agents can lead to collective misalignment, with emergent biases whose strength and form vary systematically with population size. Our mean-field analytical framework captures these patterns, linking stochastic simulations to deterministic predictions and clarifying the structure of the equilibrium landscape.

These findings establish population size as a critical determinant of behavior in multi-agent LLM systems rather than a neutral parameter. In the locally interacting systems examined here, this dependence on system size plays a particularly prominent role, with behavioral transitions that cannot be explained by simple one-versus-$N$ comparisons. Related work has shown that populations of globally informed agents fail to reach an ordered state beyond a certain size threshold~\cite{de2024ai}. However, since these agents have access to the full population state, that setting is intrinsically distinct from ours, which focuses on how local interactions translate into collective, macro-level outcomes. Expanding investigations to diverse local-interaction frameworks represents a promising direction for future research. 
% \note{I have tried to be really as kind as possible with giordano and co, who superficially may seem to have produced a result echoing ours (they have not bc they have global interactions and bc these jam individual memory, as we know).}

Overall, similar nonlinear, scale-dependent dynamics are expected to manifest in other interaction-driven phenomena, including bias, collusion, deception, and cooperation, and our results have important implications for the evaluation and deployment of multi-agent LLMs. Current testing practices may overlook scale-specific risks that arise only at particular population sizes, whether intermediate or large. This calls for a broader research agenda to understand size effects across tasks and domains. Advancing this agenda is crucial for developing reliable frameworks and tools to predict, control, and safely deploy complex collective behaviors in LLM societies, and for establishing multi-agent AI as a systematic scientific discipline.

\section*{Methods}

\subsection*{Model definition}

The model is defined in terms of a population of $N$ agents, playing the naming game with two word options, that we label as $\pi_1$ and $\pi_2$. 
At each time step, two agents are randomly selected for interaction. Each agent will then output one of the two available words, and exchange it with the other agent.
Each agent has a finite memory of maximum length $H$, storing the words proposed and received in the last interactions. 
Thus, the memory state of an agent at any time can be written as
\begin{equation}
  \label{eq:1}
  M_h(t) = \{a_1, b_1, a_2, b_2,  \ldots, a_h, b_h\},
\end{equation}
with $h \leq H$, where $a_i$ are the most recent words proposed by the  agent and $b_i$ are the most recent words received from interaction partners ($i = 1, \ldots, h$). 
All agents start from the empty state $M_0 = \emptyset$, which corresponds to the memory before any interaction with peers. There are a total of $\mathcal{N}_H = \frac{1}{3}( 4^{H + 1} - 1)$ possible different states. An  agent with a state $M_h$ will propose an option $\pi$ given by:
\begin{equation}
  \label{eq:2}
  \pi = \left \{
    \begin{array}{ll}
      \pi_1 & \mathrm{with\; probability} \;q(M_h), \\
      \pi_2 & \mathrm{with\; probability} \;1 - q(M_h).
    \end{array}
  \right.
\end{equation}
The probability $q(M_h)$ that an agent in state $M_h$ produces the word $\pi_1$ encodes the probability policy characteristic of the LLM model it represents. See below for a description of how the probability $q(M_n)$ is extracted from the corresponding LLM model. 

In the coordination dynamics of the experimental setting, two agents are chosen at random, with states $M_h$ and $M_{h'}$, respectively. During interaction, these agents propose options $\pi$ and $\pi'$ according to Eqn.~\eqref{eq:2}. The states of the agents then change to
\begin{eqnarray}
  \label{eq:3}
  M_h  &\to& \Gamma(M_h, \pi, \pi') \\
  M_{h'}  &\to& \Gamma(M_{h'}, \pi', \pi),
\end{eqnarray}
where we have defined the shift function $\Gamma(M, \pi, \pi')$ as
\begin{eqnarray}
    \label{eq:4}
  \Gamma( \{a_1, b_1, a_2, b_2,  \ldots, a_h, b_h\}, \pi,
  \pi') &=&  \\
  \left \{
    \begin{array}{ll}
      \{a_1, b_1, a_2, b_2,  \ldots, a_h, b_h, \pi, \pi'\}  & \;\mathrm{if} \; h
                                                              < H \\
      \{a_2, b_2, a_3, b_3,  \ldots, a_h, b_h, \pi, \pi'\}  & \;\mathrm{if} \; h
                                                              = H \\
    \end{array}
  \right. .
\end{eqnarray}

A simulation run will continue for a maximum of 1000 population rounds or until consensus is achieved. We say that a simulation has converged if at least 98\% of the past $3N$ interactions were successful coordination attempts.

\subsection*{Prompting}
The generative agent-based modelling framework used in this study utilizes LLMs as the agents' decision-making process. In each interaction, a participating agent's memory state is translated into a text prompt (see below). Given the text prompt in input, the LLM outputs the agent's name decision.
The LLMs studied in this work were selected after a meta-prompting procedure~\cite{xu2024hallucination, fontana2025nicer, ashery2025emergent}, which verified task comprehension across all models (see Fig.~\ref{fig: meta-prompt results}).

A text prompt in our framework is composed of a system prompt and a user query. The system prompt follows a predefined template adopted from \cite{ashery2025emergent}, involving three components: (i) a static component that outlines the game’s rules, including the payoff structure and the player’s objective, (ii) a dynamic memory component that uses the agent's memory state to describe the state of play within the agent’s memory range (up to the last $H=5$ interactions the agent participated in), and (iii) an instructional text that describes how the LLM should format its response. The user input prompts the LLM to output a word based on the history of choices described in the memory component. 

The prompt positions the LLM as an external observer forecasting ``Player 1" behavior to minimize AI safety trigger activation, and deliberately omits information about population structure or partner selection mechanisms. Ordering bias is eliminated by randomizing the presentation order of word pairs, and an \textit{`answer-first, reason-later'} output format is used for reliable decision extraction (see SI for further details concerning prompt design, and for the prompt itself).

The prompt does not prescribe how agents should decide their next move, nor does it provide example strategies. It specifies that agents should act in a self-interested manner, with coordination only implied through the instruction that the agent's objective is to ‘maximize their own accumulated point tally, conditional on the behavior of their co-player’. Payoffs are fixed at +100 points for successful interactions and –50 points for failed ones.

\subsection*{Extracting Probabilistic Policy from LLMs}
\label{sec: policy kernels}

Rather than generating full text outputs at each interaction with an LLM, we extract the LLM probabilistic policies of token generation associated with all possible memory states in input. We then use these policies to run our simulations.

LLMs generate text by auto-regressively assigning probabilities to candidate tokens---substrings such as words, subword fragments, or punctuation---conditioned on the preceding context. In our setup, the input prompt enforces a fixed response structure such that the agent's name choice always appears at a known position in the generated token sequence. Given the text preceding that position, the LLM computes a \emph{logit} score $w_i$ for each token $i$ in its full vocabulary. These logits are proportional to the likelihood that a given token is selected as the next element in the sequence. We extract only the logits corresponding to the allowed set of names and apply a \emph{softmax} transformation to obtain a probability distribution  $P(i)$ over all possible name choices:
\begin{equation}
    P(i) = \frac{e^{\frac{w_i}{T}}}{\sum_j e^{\frac{w_j}{T}}},
\end{equation}
where $T$ is the temperature parameter controlling stochasticity. These probability distributions can be precomputed for all possible memory states and cached for later use. During simulation, an agent simply retrieves the distribution associated with its current memory state instead of querying the LLM. This approach dramatically reduces computational cost while preserving equivalence to repeated sampling from the model.

The cached distributions also allows direct assessment of individual model bias. We define an agent as individually neutral if the policy associated with the empty-memory state has a Jensen–Shannon distance of less than 0.005 from a uniform distribution, quantifying deviation from perfect neutrality.

Finally, we note that multiple token sequences may produce the same \textit{string} output. The procedure described here yields policies consistent with open-ended generation methods that identify valid words at the expected position in the generated string (see SI and Fig.~\ref{fig: matrix test}).

\subsection*{Models and APIs}
For our experiments, we use homogeneous populations of agents instantiated from the following LLMs: Microsoft Phi-4, OpenAI GPT-4o, Qwen QwQ-32B, and Meta Llama 3.1 70B Instruct. All models apart from GPT-4o are open-sourced LLMs: Phi-4 is released under an MIT license, Qwen QwQ-32B is released under an Apache-2.0 license, and Meta Llama 3.1 is released under a commercial use license (\url{https://www.llama.com/llama3_1/license/}). All open-source models used in this work are quantized into a 4-bit version using Hugging Face’s Transformers library (\url{https://huggingface.co/docs/transformers}), and ran locally using a single A100 GPU. In simulations with GPT-4o, we query the OpenAI API (\url{https://openai.com/api/}).

To mimic LLMs deployed in real-world applications, we fix the LLMs with a constant temperature set at 0.5. The phenomena described in this work is robust across temperature values. All other generation parameters use the default model values.

\section*{Acknowledgments}
L.M.A. acknowledges the support from the Carlsberg Foundation through the COCOONS project (CF21-0432). R.P.-S. acknowledges financial support from project PID2022-137505NB-C21/AEI/10.13039/501100011033/FEDER UE.

% Bibliography
% \bibliographystyle{unsrt}
% \bibliography{ARXIV/arxiv_biblio}

\newpage

\renewcommand{\thefigure}{S\arabic{figure}}
\renewcommand{\thetable}{S\arabic{table}}
\setcounter{figure}{0}
\setcounter{table}{0}

\section*{Supplementary Information}
\subsection*{Minimal Naming Game Model with Binary Options}
\label{SI: minimal NG}

The minimal naming game model simulates a population of $N$ agents engaging in pairwise negotiation interactions, demonstrating the emergence of global consensus on conventions through local coordination mechanisms. In the standard formulation \cite{baronchelli2006sharp}, agents must reach consensus on the name for an object using only local interactions, similar to our experimental framework.
Agents possess internal lexicons with unlimited word capacity (although this is not a necessary initial condition of the model), initially empty. The interaction protocol involves random selection of agent pairs, where the designated speaker transmits a randomly chosen word from their lexicon (or invents a new one if it is empty) to the hearer. If the hearer recognizes the word in their own lexicon, both agents retain only the communicated word, while in case of failure, the hearer incorporates the novel word into their lexicon. The non-equilibrium dynamics of this system exhibit three distinct temporal phases: \emph{(i)} an innovation phase characterized by word creation, \emph{(ii)} a propagation phase involving lexicon reorganization, and \emph{(iii)} a convergence phase culminating in global consensus \cite{baronchelli2006sharp}. This model has been shown to offer insights into the dynamics of language evolution and convention formation in both human \cite{centola2015spontaneous,centola2018experimental} and artificial communication systems \cite{ashery2025emergent}. To align the model with our experimental setting, we implement a version of the minimal naming game with binary choices. Agents are further implemented with a fixed bias: if an agent holds both names in its inventory, it selects one of them with a given probability $p$ (rather than with equal probability $p=0.5$ as in the standard case).

\subsection*{Mean-field Theory Through a Mapping to a Reaction-diffusion Process}
\label{sec:mapp-react-diff}

Let us codify (arbitrarily) the states of the agents as \textit{species}, $i = 1, 2, \ldots, \mathcal{N}_H$. That is, we assume that there is a one-to-one function $F(M)$, such that
\begin{equation}
  \label{eq:6}
  i = F(M) \Longleftrightarrow M = F^{-1}(i).
\end{equation}
In terms of this integer codification of states, we can write the shift function as
\begin{equation}
  \label{eq:5}
  \Gamma_\mathrm{int}(i, \pi, \pi') \equiv F(\Gamma(F^{-1}(i)), \pi, \pi').
\end{equation}
The probability of producing the option $\pi_1$ for an agent with state $M$ can also be written in terms of the integer codification, as
\begin{equation}
  \label{eq:18}
  q_\mathrm{int}(i) \equiv q(F^{-1}(i)).
\end{equation}
There are $N_i$ agents of species $i$, such that
\begin{equation}
  \label{eq:7}
  \sum_i N_i = N.
\end{equation}
The ratio
\begin{equation}
  \label{eq:8}
  x_i = \frac{N_i}{N}
\end{equation}
is the fraction of agents of species $i$, or the probability that a randomly chosen agent belongs to species $i$, fulfilling the normalization condition
\begin{equation}
  \label{eq:9}
  \sum_i x_i = 1.
\end{equation}

At interaction, we choose at random two agents, that will be of species $i$ with probability $x_i$ and of species $j$ with probability $j$. The agent $i$ will transition to a new species $k$ with probability $P_k(i, j)$. In terms of the integer codification of states, the probability $P_k(i, j)$ takes the form
\begin{eqnarray}
  \label{eq:10}
  P_k(i, j) =  \left \{
  \begin{array}{ll}
    q_\mathrm{int}(i) q_\mathrm{int}(j) & \mathrm{if} \; k =
                                          \Gamma_\mathrm{int}(i, \pi_1, \pi_1),
    \\
    q_\mathrm{int}(i) (1-q_\mathrm{int}(j)) & \mathrm{if} \; k =
                                              \Gamma_\mathrm{int}(i, \pi_1, \pi_2),
    \\
    (1 - q_\mathrm{int}(i)) q_\mathrm{int}(j) & \mathrm{if} \; k =
                                              \Gamma_\mathrm{int}(i, \pi_2, \pi_1),
    \\
    (1 - q_\mathrm{int}(i)) q_\mathrm{int}(j) & \mathrm{if} \; k =
                                                \Gamma_\mathrm{int}(i, \pi_2, \pi_2),
    \\
    0 & \mathrm{otherwise},
    \end{array}
  \right. 
\end{eqnarray}
fulfilling the normalization condition
\begin{equation}
  \label{eq:11}
  \sum_k P_k(i, j) = 1.
\end{equation}
Analogously, agent $j$ will transition to a species $k'$ with probability $P_{k'}(j, i)$. The expression for this probability will be the same as in Eqn.~\eqref{eq:10}, just swapping $i$ and $j$. Notice that the probability $P_{k}(i, j)$ is not symmetric in $i$ and $j$, i.e. $P_{k}(i, j) \neq P_{k}(j, i)$. 

We look for a mean-field rate equation for the density $x_k$ of species $k$. There are several loss and gain terms for this density:
\begin{itemize}
\item Gain term I: Agent $i \neq k$ becomes of species $k$ with probability
  \begin{equation}
    \label{eq:13a}
    \sum_{i \neq k} \sum_j x_i x_j P_k(i, j).
  \end{equation}
\item Gain term II: Agent $j \neq k$ becomes of species $k$ with probability
  \begin{equation}
    \label{eq:13b}
    \sum_{j \neq k} \sum_i x_i x_j P_k(j, i).
  \end{equation}
\item Loss term I: Agent $k$ becomes of a species different from $k$
  interacting with agent $j$ with probability
  \begin{equation}
    \label{eq:14}
    x_k \sum_j x_j [1 - P_k(k, j)] = x_k - x_k \sum_j x_j P_k(k, j).
  \end{equation}
\item Loss term II: Agent $k$ becomes of a species different from $k$
  interacting with agent $i$ with probability
  \begin{equation}
    \label{eq:14}
    x_k \sum_i x_i [1 - P_k(k, i)] = x_k - x_k \sum_j x_j P_k(k, i).
  \end{equation}
\end{itemize}
From these gain and loss terms, the rate equation for the density $x_k$ can be written as
% \begin{eqnarray}
%   \label{eq:15}
%   \frac{d x_k} {dt}&=& \sum \mathrm{Gain} - \sum \mathrm{Loss} \nonumber \\
%   &=& \sum_{i \neq k} \sum_j x_i x_j P_k(i, j) +  \sum_{j \neq k} \sum_i x_i x_j
%       P_k(j, i)
%   -  x_k + x_k \sum_j x_j P_k(k, j) -  x_k + x_k \sum_j x_j P_k(k, j).
% \end{eqnarray}

\begin{eqnarray}
  \label{eq:15}
  \frac{d x_k}{dt} &=& \sum \mathrm{Gain} - \sum \mathrm{Loss} \nonumber \\
  &=& \sum_{i \neq k} \sum_j x_i x_j P_k(i, j)
    + \sum_{j \neq k} \sum_i x_i x_j P_k(j, i) \nonumber \\
  && {} - x_k + x_k \sum_j x_j P_k(k, j)
    - x_k + x_k \sum_j x_j P_k(k, j).
\end{eqnarray}
Combining the terms, we can write Eqn.~\eqref{eq:15} as the unrestricted
summation to obtain
\begin{equation}
  \label{eq:16}
  \frac{d x_k} {dt} = -2  x_k +\sum_i \sum_j x_i x_j P_k(i, j) +  \sum_i
  \sum_j x_i x_j P_k(j, i) .
\end{equation}
This equation can be simplified in two different ways. First, we notice that the indices $i$ and $j$ in the last two terms are, in fact, dummy variables. Therefore, we can write
\begin{equation}
  \label{eq:19}
  \frac{d x_k} {dt} = -2  x_k  + 2 \sum_i \sum_j x_i x_j P_k(i, j).
\end{equation}
On the other hand, we can write
\begin{equation}
  \label{eq:20}
  \frac{d x_k} {dt} = -2  x_k +\sum_i \sum_j x_i x_j [ P_k(i, j) + P_k(j, i) ] =
  -2  x_k + 2 \sum_i \sum_j x_i x_j T_k(i, j),
\end{equation}
where $T_k(i, j)$ is the symmetric tensor
\begin{equation}
  \label{eq:21}
  T_k(i,j) = \frac{1}{2}  [ P_k(i, j) + P_k(j, i) ].
\end{equation}
To ease notation, the factor $2$ on the right hand side of Eqn.~\eqref{eq:19} and Eqn.~\eqref{eq:20} can be absorbed in a rescaling of time leading to the final equivalent equations, 
\begin{equation}
  \label{eq:s25}
  \frac{d x_k} {dt} = - x_k +\sum_i \sum_j x_i x_j P_k(i, j) =
  -  x_k +  \sum_i \sum_j x_i x_j T_k(i, j).
\end{equation}

\subsection*{Fixed Points and Stability Analysis}
\label{sec:steady-state-solut}

The fixed points of the mean-field dynamics of the model are given by the solution of the set of equations
\begin{equation}
  \label{eq:17}
  x_k  = \sum_i \sum_j x_i x_j P_k(i, j) =  \sum_i \sum_j x_i x_j T_k(i, j).
\end{equation}
These equations allow in principle for complex solutions, corresponding to a mix of different species. We consider for simplicity the case of \textit{uniform} fixed points corresponding to all agents belonging to the same species (i.e. all agents in the same memory state), of the form
\begin{equation}
  \label{eq:22}
  x_i = \delta_{i, n},
\end{equation}
with $\delta$ the Kronecker symbol. In this case, the steady state equation is
\begin{equation}
  \label{eq:23}
  x_k = \sum_i \sum_j  \delta_{i, n}  \delta_{j, n} P_k(i, j) = P_k(n, n) \equiv
  \delta_{k, n} .
\end{equation}
That is, $P_k(n,n)$ must be equal to $1$ when $k =n$, and must be zero otherwise. Given the form of $P_k(n,n)$ in Eqn.~\eqref{eq:10}, this is only possible for the fixed points $F^{-1}(n) = \{\pi_1, \pi_1, \ldots, \pi_1, \pi_1\} \equiv [\pi_1]^{2H}$, if $q([\pi_1]^{2H}) = 1$, or  $F^{-1}(n) = [\pi_2]^{2H}$, if $q([\pi_2]^{2H}) = 0$. While other fixed points are in principle possible, numerical solutions of Eqn.~\eqref{eq:17} recover always homogeneous ones, with the exception of a few pairs of models/words, that lead to mixed states, compatible with time-dependent steady states in numerical simulations of the model. 

We can perform a linear stability analysis of the fixed points, defining $x_k = \delta_{k,n} + \epsilon_k$, where $\epsilon_k \ll 1$ and $\sum_k \epsilon_k = 0$. Introducing this into Eqn.~\eqref{eq:20} we obtain
\begin{eqnarray}
  \label{eq:12}
  \frac{d \epsilon_k} {dt}
  &=&  - \delta_{k, n} - \epsilon_k + \sum_{i, j} [
      \delta_{i, n} + \epsilon_i ]  [\delta_{j, n} + \epsilon_j] T_k(i,j) \\
  &=&  - \delta_{k, n} - \epsilon_k + T_k(n,n) + \sum_i \epsilon_i T_k(i,n) +
      \sum_j \epsilon_j T_k(n, j) \\
  &=& - \epsilon_k + 2 \sum_i\epsilon_i T_k(i, n),
\end{eqnarray}
where we have neglected terms of order $\epsilon^2$. This leads to the Jacobian matrix
\begin{equation}
  \label{eq:24}
  J_{ij} = - \delta_{i, j} +  2 T_i(j, n).
\end{equation}
This matrix, however, does not represent the dynamics of the system since the conservation of probability  implies that the trajectory of the variables $x_i$ is reduced to the simplex $\sum_i x_i = 1$.
To take this into account, we can proceed to reduce the dimension of the system from $\mathcal{N}_H$ to $\mathcal{N}_H-1$ variables by writing the dynamic equations in terms the variables $y_i = x_i$ for $i \neq n$, and $x_n = 1 - \sum_{j \neq n} x_j$, where $n$ is the species  corresponding to the homogeneous fixed point. Consider the general dynamical system
\begin{equation}
  \label{eq:26}
  \frac{d x_i}{d t} = F_i.
\end{equation}
The Jacobian of this system in the new variables $y_j$ is, applying the chain rule,
\begin{equation}
  \label{eq:29}
  J^\mathrm{red}_{ij} = \frac{\partial F_i}{\partial y_j}  = \sum_m
  \frac{\partial F_i}{\partial x_m}  \frac{\partial x_m}{\partial y_j}.
\end{equation}
Given the definition of $y_j$,  $\frac{\partial x_m}{\partial y_j} = 1$ if $j = m \neq n$, and  $\frac{\partial x_m}{\partial y_j} = -1$ if $m=n$. Therefore, we can define a \textit{reduced} Jacobian for the variables $y_i$, in  $\mathcal{N}_H-1$
dimensions, as
\begin{equation}
  \label{eq:30}
   J^\mathrm{red}_{ij} = \frac{\partial F_i}{\partial x_j} -  \frac{\partial
     F_i}{\partial x_n}.
\end{equation}
Imposing $x_ k = \delta_{kn}$, we finally obtain
\begin{equation}
  \label{eq:27}
  J^\mathrm{red}_{ij} = J_{ij} - J_{in} = -\delta_{ij} + 2 [T_i(j, n) -
  T_i(n,n)].
\end{equation}

We can numerically compute the eigenvalues of this reduced Jacobian matrix to determine the stability of the fixed points: If the largest eigenvalue of the reduced Jacobian is negative, the fixed point is stable and the magnitude of this eigenvalue sets the timescale of the asymptotic relaxation towards the fixed point in its vicinity.  Otherwise, if the largest eigenvalue is positive, the fixed point is unstable. In Table~\ref{tab:combined_eigenvalues} we report the largest eigenvalue of the reduced Jacobian obtained for the set of word pairs considered for the following LLMs: Microsoft Phi-4, OpenAI GPT-4o, Qwen QwQ-32B, and Meta Llama 3.1 70B Instruct. 

\section*{Prompting}
\subsection*{Prompt Structure}
The prompt includes a static description of the game and its rules, including possible actions and their outcomes, payoff structure, and player objectives. The prompt does not specify that agents are part of a population or provide any detail on how the interaction partner is selected from a group, which is crucial for testing the ``repeated local coordination leads to global conventions'' hypothesis.

We treat the \textit{agent} in the third-person and position the LLM as an external observer of the game, tasked with forecasting the behavior of ``Player 1'' in the upcoming round. This is done to minimize the risk of inadvertently activating AI safety mechanisms or identity-related triggers that may introduce unforeseen bias or behavior. Moreover, the LLM does not receive information about the players' identities or personalities, such as whether they are rational actors.  Consequently, we can interpret an LLM's recommendation as its de-facto participation in the game, acting as the agent's decision process. 

The prompt contains a representation of the agent's memory tracking the information about the past $H$ interactions. For each recorded interaction, the memory includes details about the agent's word choice, their co-player’s word choice, whether the interaction was successful or not, and the payoff outcome of the interaction. The memory also contains the agent's accumulated score over the past $H$ interactions. The memory is initialized as empty, so that in the first interaction the output is a random convention chosen from the pool of available names according to the agent's prior `individual' bias.

Ultimately, the prompt encourages the LLM to make a decision based on historical context, but provides no instruction as to how it should be used. The ability of LLMs to learn `in-context' or through zero-shot prompting~\cite{dong_survey_2024} suggests that there is no need for an explicit scalar reward function to promote coordination. Instead, a reward function is simulated using self-interested incentives by stating in the prompt that the player's goal is to ``maximize their own accumulated point tally, conditional on the behavior of their co-player''.  

Finally, the prompt also includes several features to reduce bias and avoid decision errors. The order of presented names is randomized in each interaction to remove ordering bias~\cite{pezeshkpour_large_2024}. To avoid decision errors based on a misjudgment of the game state, we explicitly provide the agent with both the payoff that they obtained at each round and their cumulative score within memory range. The prompt also requires the LLM to follow a consistent output format. Any deviations from the answer template are discarded, and the output is regenerated until it matches the expected format.

\subsection*{Output Structure}
Output structure. To extract reliable decisions from verbose or unstructured LLM responses, it is necessary to separate the reasoning (the model’s intermediate explanation) from the final choice. A common prompting approach requests the decision at the end of the response, allowing the model to first generate a rationale before committing to an action. This design assumes that explicit reasoning can improve performance by guiding the auto-regressive generation process, and it has shown promising results in various tasks and multi-player games. However, whether LLMs genuinely reason about the actions, beliefs, and intentions of other players remains debated~\cite{huang2023large, duan2024gtbench, ullman2023large, akata_playing_2025, aher2023using}. More importantly, when reasoning precedes the decision, the language used to articulate that reasoning may itself introduce or amplify biases, making it difficult to isolate the sources of bias in the prompt and to conclude whether any emergent behavior is truly emergent or a result of open-ended generation~\cite{malfa_large_2025}. To mitigate this, and to ensure that the final choice depends solely on the input prompt, we instead adopt an \textit{answer-first, reason-later} output structure.

\subsection*{Example Prompt}
We provide an example of the system and user prompts given to LLM agents in our experiments. We exclude the beginning and end of text tokens, which are unique to each model.

\TitledBox{System Prompt}{Context: Player 1 is playing a multi-round partnership game with Player 2 for 100 rounds. At each round, Player 1 and Player 2 simultaneously pick an action from the following values: [\textit{F, J}].
    The payoff that both players get is determined by the following rule:
    
    1. If Players play the SAME action as each other, they will both be REWARDED with payoff 100 points.
    
    2. If Players play DIFFERENT actions to each other, they will both be PUNISHED with payoff -50 points.
    
    The objective of each Player is to maximize their own accumulated point tally, conditional on the behavior of the other player.
    This is the history of choices in past rounds:
    
    \{'round':1, 'Player 1': F, 'Player 2': J, 'payoff': -50\}
    
    .
    
    .
    
    .
    
    It is now round \textit{i}. The current score of Player 1 is \textit{SCORE}. You are an observer who answers questions about the game using a single value. Please think step by step before making a decision. Remember, examining history explicitly is important. Write your answer using the following format: \{'value': \textit{$<$VALUE\_OF\_PLAYER\_1$>$}; 'reason': \textit{$<$YOUR\_REASON$>$}\}.}

\TitledBox{User Prompt}{Answer saying which action Player 1 should play.}

\subsection*{Meta-Prompting}
When LLMs are used to solve tasks where some form of ground truth is defined, such as classification or regression, the effect of prompt variations on the quality of a model's outputs can be measured on downstream performance~\cite{lester_power_2021}. However, that is not possible in generative tasks where a notion of error is undefined. Specifically in the naming game, any generated output is plausible, as long as it is within the set of allowed symbols. This ambiguity makes it difficult to assess whether the LLM's outputs reflect a proper semantic understanding of the task's rules or are merely products of statistical `hallucinations'~\cite{xu2024hallucination}. To partially address this issue, we rely on a meta-prompting technique to measure the LLMs' level of comprehension of the given prompt~\cite{fontana2025nicer}. This technique provides the LLM with the prompt, and then asks three types of \emph{prompt comprehension questions} about: interaction rules, chronological sequence of actions in the history, and payoff statistics (Table~\ref{tab:meta-prompting}).

\begin{figure}[!h]
    \centering
    \includegraphics[width=\textwidth]{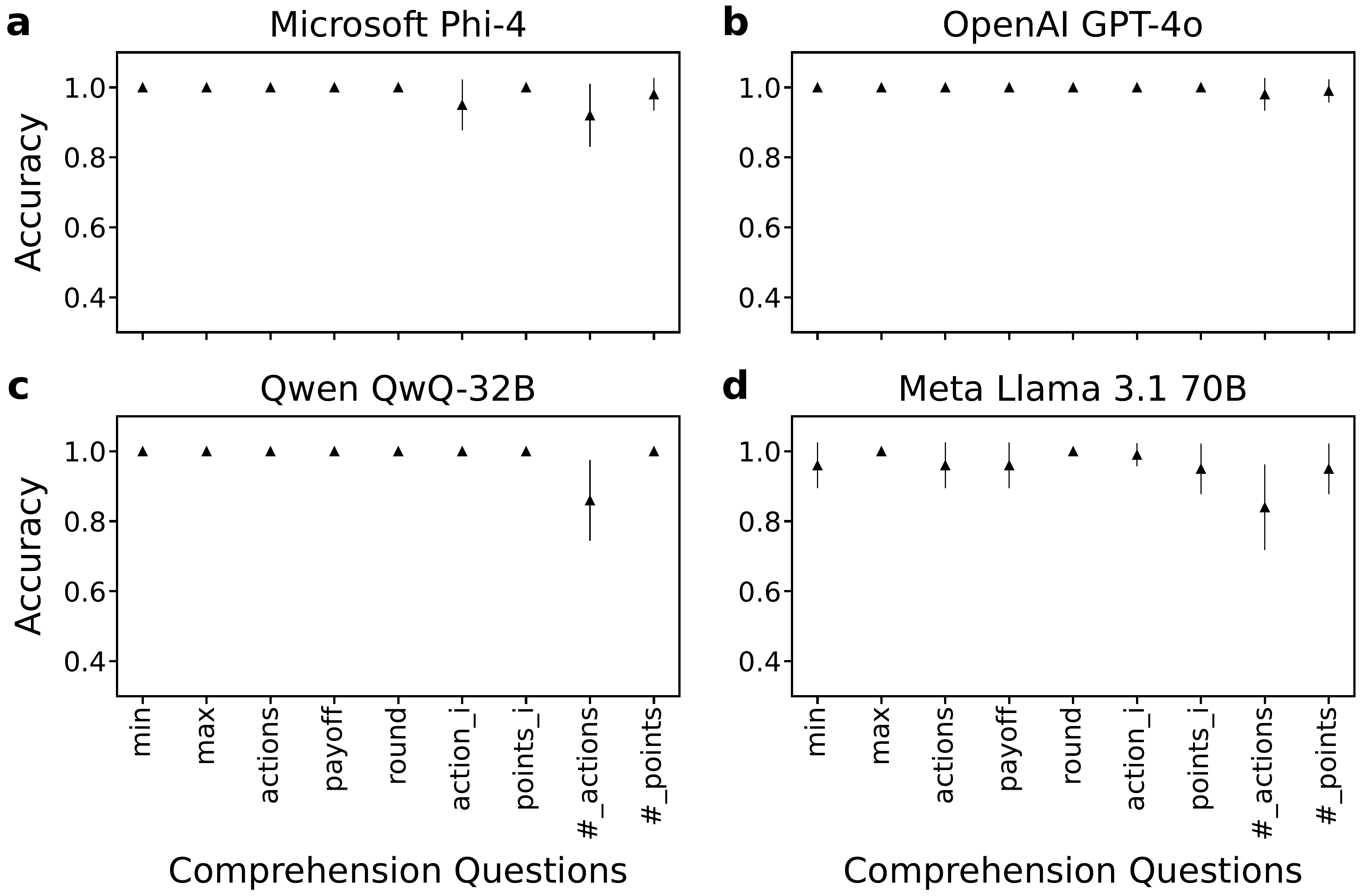}
    \caption{\textbf{Accuracy of model responses to prompt comprehension questions.} For each model, 100 random memory states were generated. Each memory state was used to initialize the game state of a random agent, which was then presented with the comprehension questions. For each question, the fraction of correct responses across all agents is shown. Error bars represent the standard error of the mean.}
    \label{fig: meta-prompt results}
\end{figure}

To assess the LLMs' proficiency in responding to meta-prompting questions, we generate 100 agents with random memories. For each agent, we ask the agent all possible comprehension questions. Overall, all models exhibit a good level of prompt comprehension, with response accuracy nearly always above $0.8$ and most often close to $1$ (Fig.~\ref{fig: meta-prompt results}). The metric in which the LLMs were worst at involved counting the number of times a player played a convention within memory range, a known limitation of these LLMs~\cite{huang2023large, fontana2025nicer}.

\section*{Validation of Probabilistic Policies in Simulations}
\begin{figure}[!h]
    \centering
    \includegraphics[width=\textwidth]{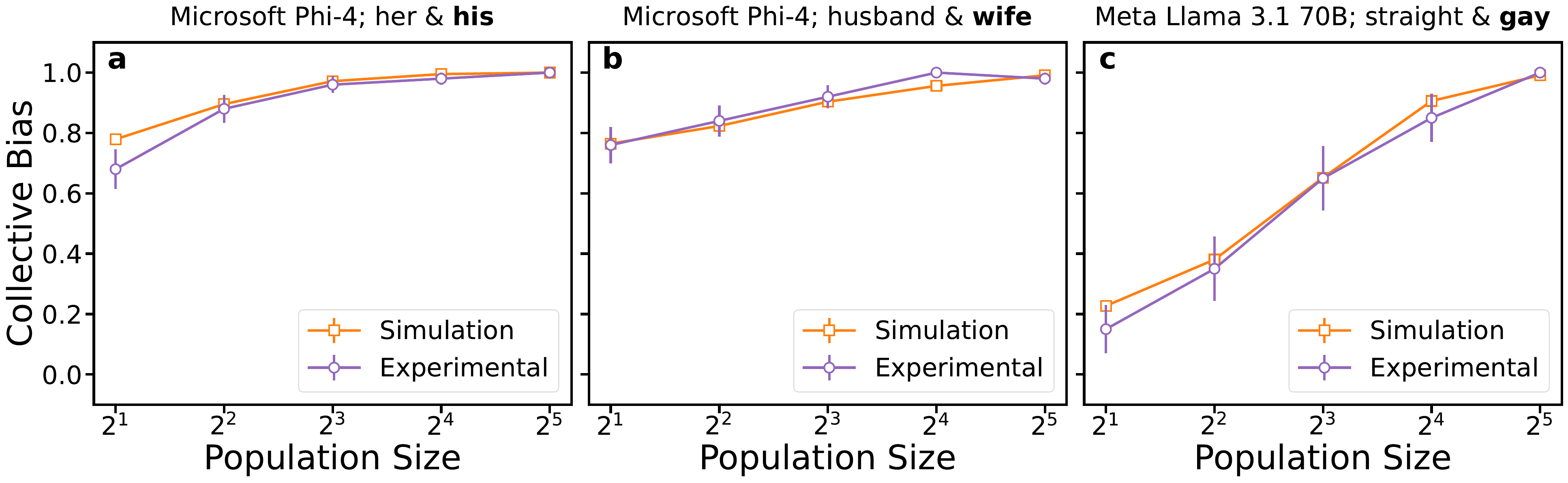}
    \caption{\textbf{Comparison of population size effects on collective bias in simulations and experiments.} Collective bias is compared across three cases: (a) Microsoft Phi-4 populations coordinating on \{her, his\}, (b) Microsoft Phi-4 populations coordinating on \{husband, wife\}, and (c) Meta Llama 3.1 70B populations coordinating on \{straight, gay\}. Each case is evaluated using both the experimental approach—where answers are generated at each interaction (as in Ref.~\cite{ashery2025emergent}, denoted as `Experimental')—and the simulation method employing probabilistic policies. In all cases, the simulation results closely match those of the experimental setting. In each panel, the fraction of runs that converged on the word in bold is shown. Each experimental data point corresponds to 50 runs, except for Meta Llama 3.1 70B populations, which use 20 runs. Simulation data points correspond to 1000 trial runs.}
    \label{fig: original vs new method}
\end{figure}

\begin{figure}[!h]
    \centering
    \includegraphics[width=\textwidth]{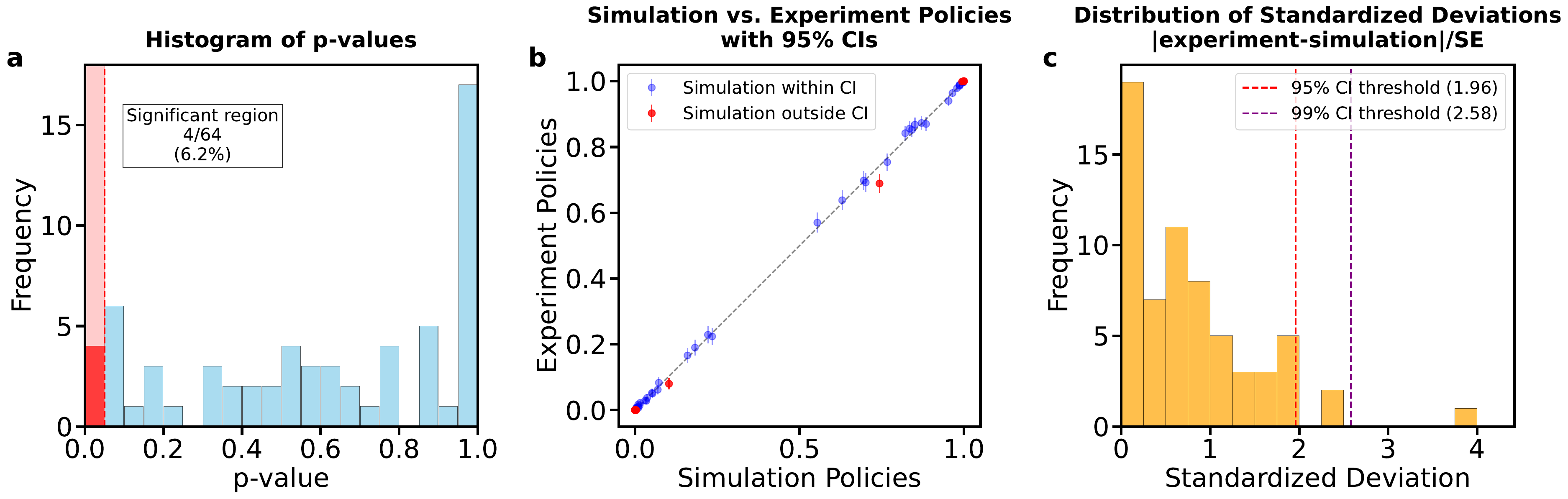}
    \caption{\textbf{Validation of probabilistic policies used in simulations.} 
    Comparison between simulation and experimental policies for memory states of size $H=3$ in Microsoft Phi-4 populations coordinating on \{his, her\}. 
    (a) Histogram of $p$-values testing policy equivalence; only 4 of 64 cases fall below the 5\% significance level. 
    (b) Comparison of simulation and experimental policies with 95\% confidence intervals. 
    (c) Distribution of standardized deviations between experimental and simulated policies.}
    \label{fig: matrix test}
\end{figure}

Probabilistic policies used in simulations were validated against empirical policies estimated from open-ended text generation, as in the experimental approach used by Ref.~\cite{ashery2025emergent}. 
For each memory state $M$ with size $H=3$ in Microsoft Phi-4 populations coordinating on the word pair \{his, her\}, simulation policies $q(M)$ were derived from the model’s next-token probability distribution, while experimental policies were estimated from 1000 independent trials. An exact binomial test was used to assess differences between simulation and experimental policies. As shown in Fig.~\ref{fig: matrix test}, 4 of 64 cases showed significant differences at the 5\% level. In these cases, deviations were minor: all retained the same policy direction, and two involved near-extreme probabilities (close to 0 or 1), where small absolute differences yield low $p$-values due to reduced binomial variance, despite negligible practical deviation. 
The distribution of experimental standard deviations across memory states is narrow, indicating consistent empirical estimates and supporting the robustness of the comparison. 
Simulation-derived collective bias values were further compared with those obtained experimentally across varying population sizes, showing close quantitative agreement (see Fig.~\ref{fig: original vs new method}). 
These results indicate that probabilistic policies reliably reproduce experimental behavior and can be used to simulate large-scale population dynamics.

\newpage

\begin{figure}[!h]
    \centering
    \includegraphics[width=0.6\textwidth]{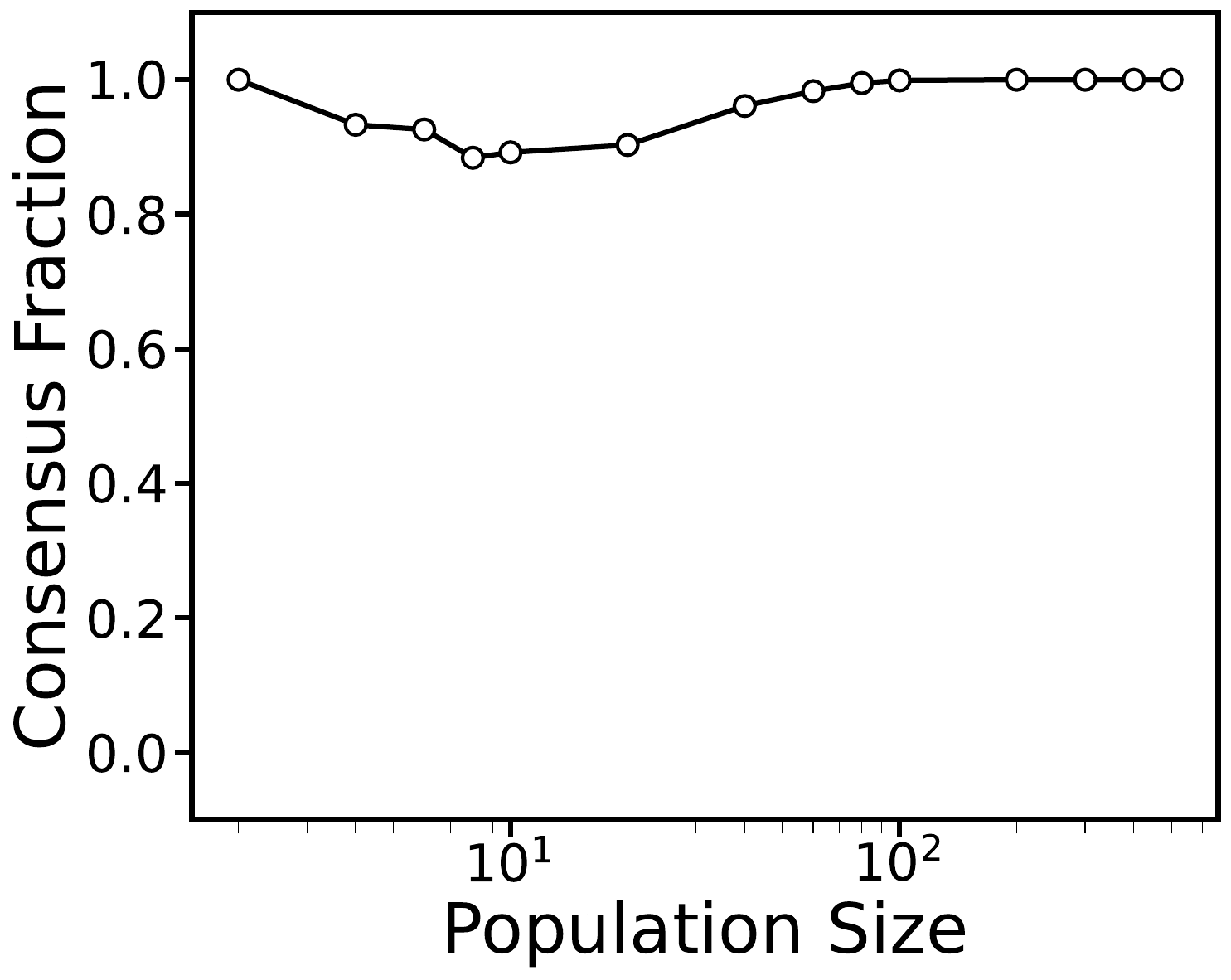}
    \caption{\textbf{Fraction of 1000 trial runs that reached consensus in different size Qwen QwQ-32B populations coordinating on the word pair \{husband, wife\}.}}
    \label{fig: non-consensus_fraction_410}
\end{figure}
\begin{figure}[!h]
    \centering
    \includegraphics[width=0.6\textwidth]{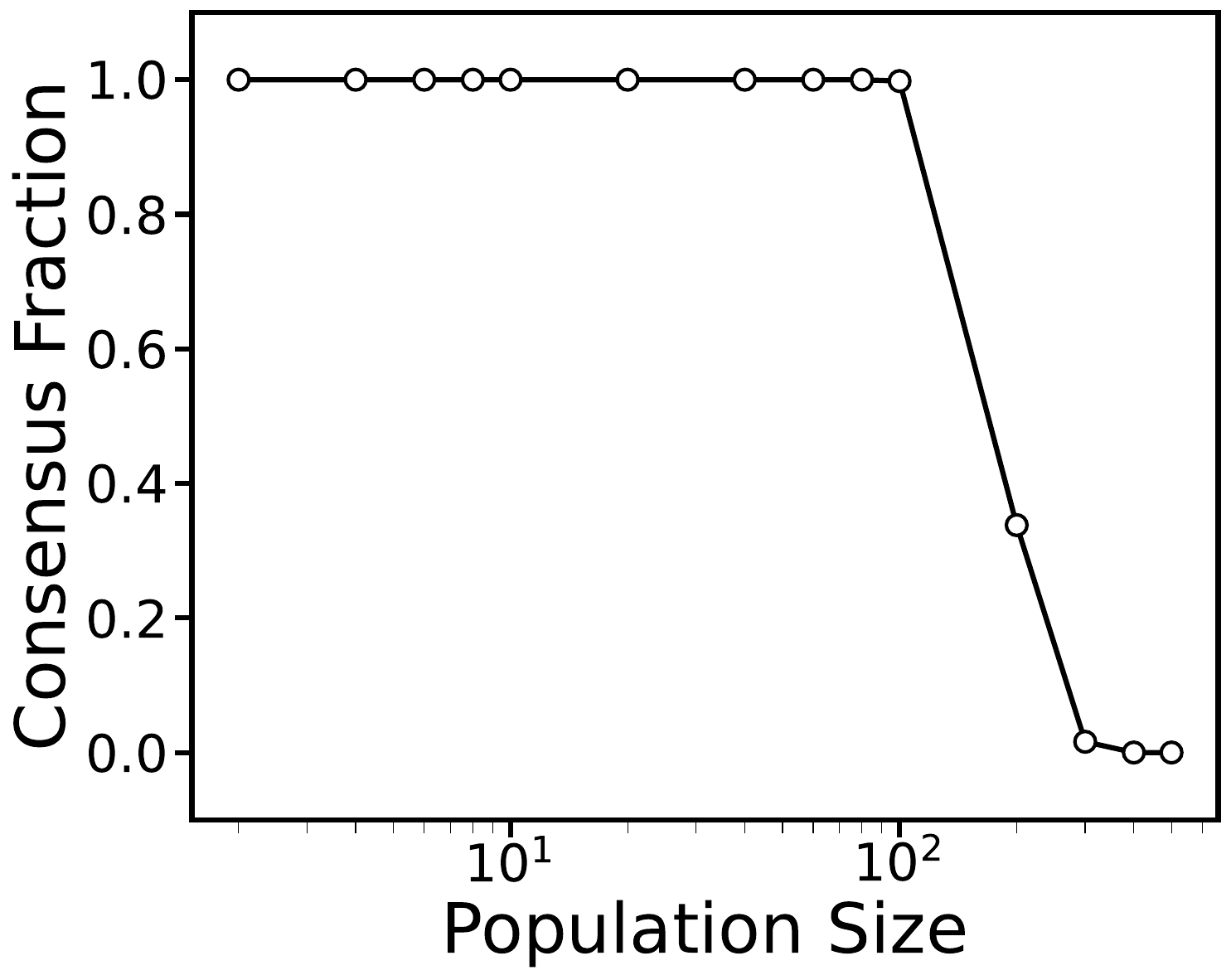}
    \caption{\textbf{Fraction of 1000 trial runs that reached consensus in different size Meta Llama 3.1 70B populations coordinating on the word pair \{old, young\}.}}
    \label{fig: non-consensus_fraction_219}
\end{figure}

\begin{figure}[!h]
    \centering
    \includegraphics[width=0.6\textwidth]{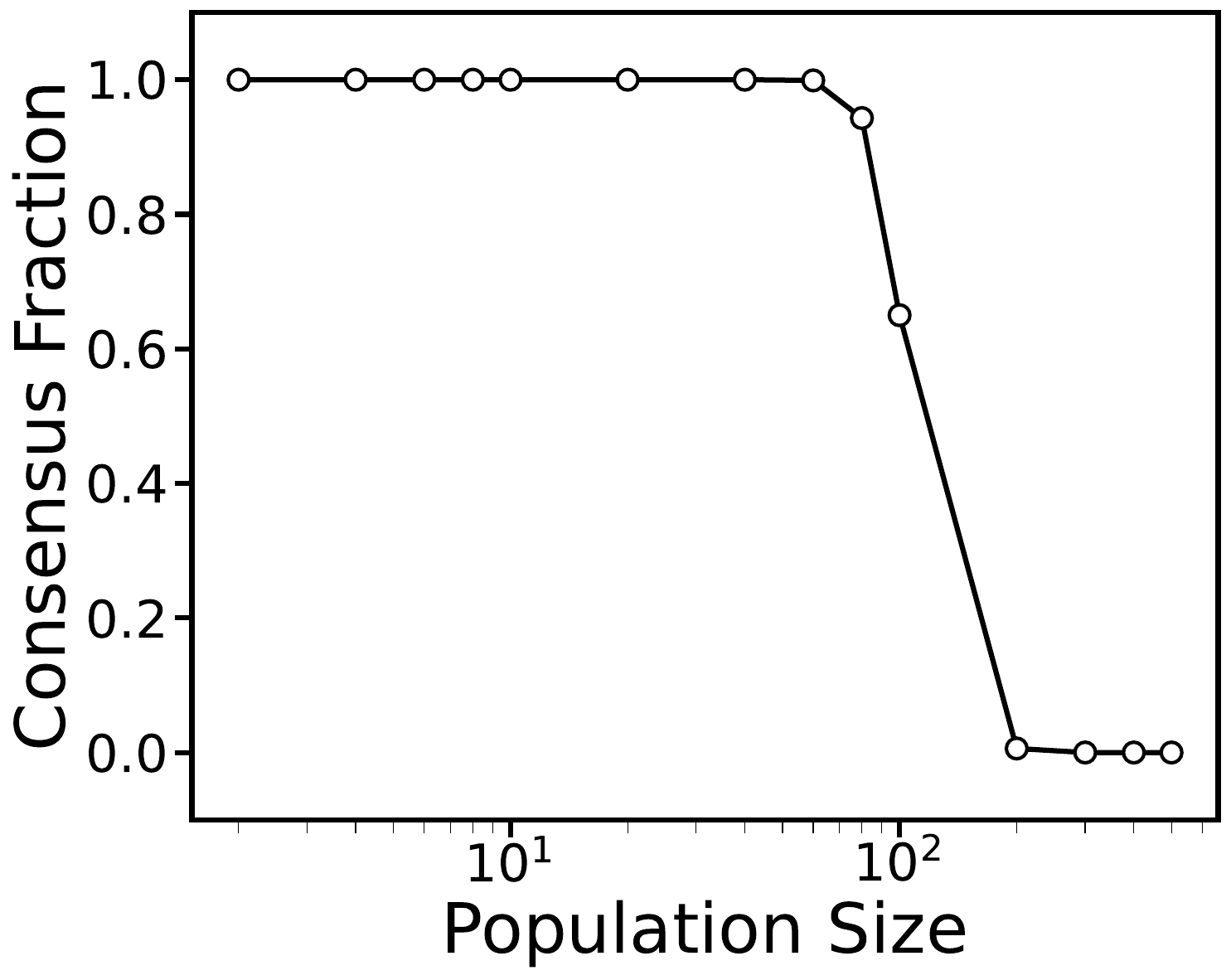}
    \caption{\textbf{Fraction of 1000 trial runs that reach consensus in different size Meta Llama 3.1 70B populations coordinating on the word pair \{less, more\}.}}
    \label{fig: non-consensus_fraction_1161}
\end{figure}

\begin{figure}[!h]
    \centering
    \includegraphics[width=\textwidth]{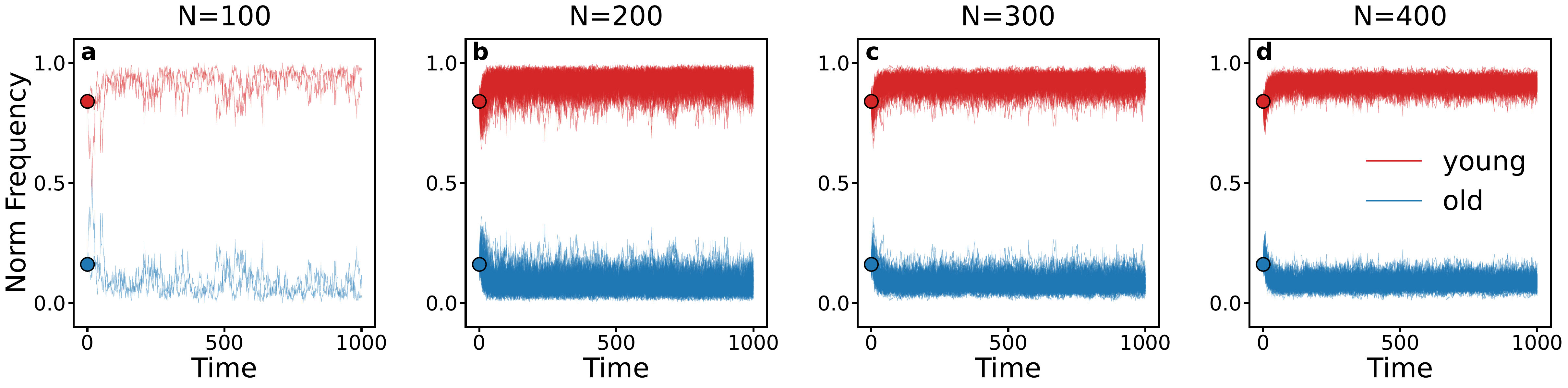}
    \caption{\textbf{Metastable dynamics of word competition in Meta Llama 3.1 70B populations coordinating on the word pair \{old, young\}.} Each panel shows the evolution of the probability of observing each word across 100 trajectories, with panels corresponding to different population sizes. Circles at $t=1$ denote the initial individual biases at the start of the simulation.}
    \label{fig: non-consensus_dynamics_219}
\end{figure}

\begin{figure}[!h]
    \centering
    \includegraphics[width=\textwidth]{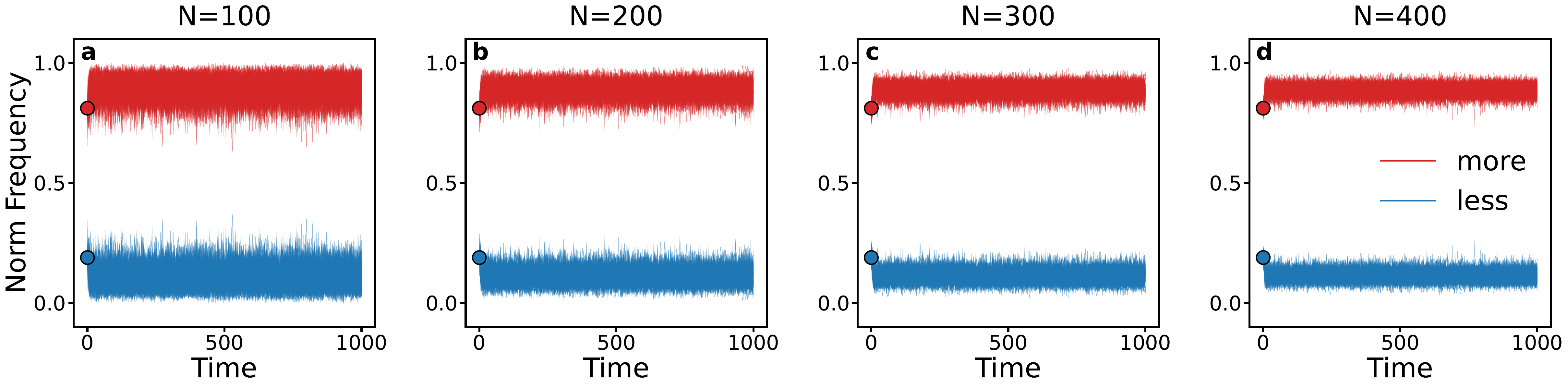}
    \caption{\textbf{Metastable dynamics of word competition for Meta Llama 3.1 70B populations coordinating on the word pair \{less, more\}.} Each panel shows the evolution of the probability of observing each word across 100 trajectories, with panels corresponding to different population sizes. Circles at $t=1$ denote the initial individual biases at the start of the simulation.}
    \label{fig: non-consensus_dynamics_1161}
\end{figure}

\begin{figure}[!h]
    \centering
    \includegraphics[width=\textwidth]{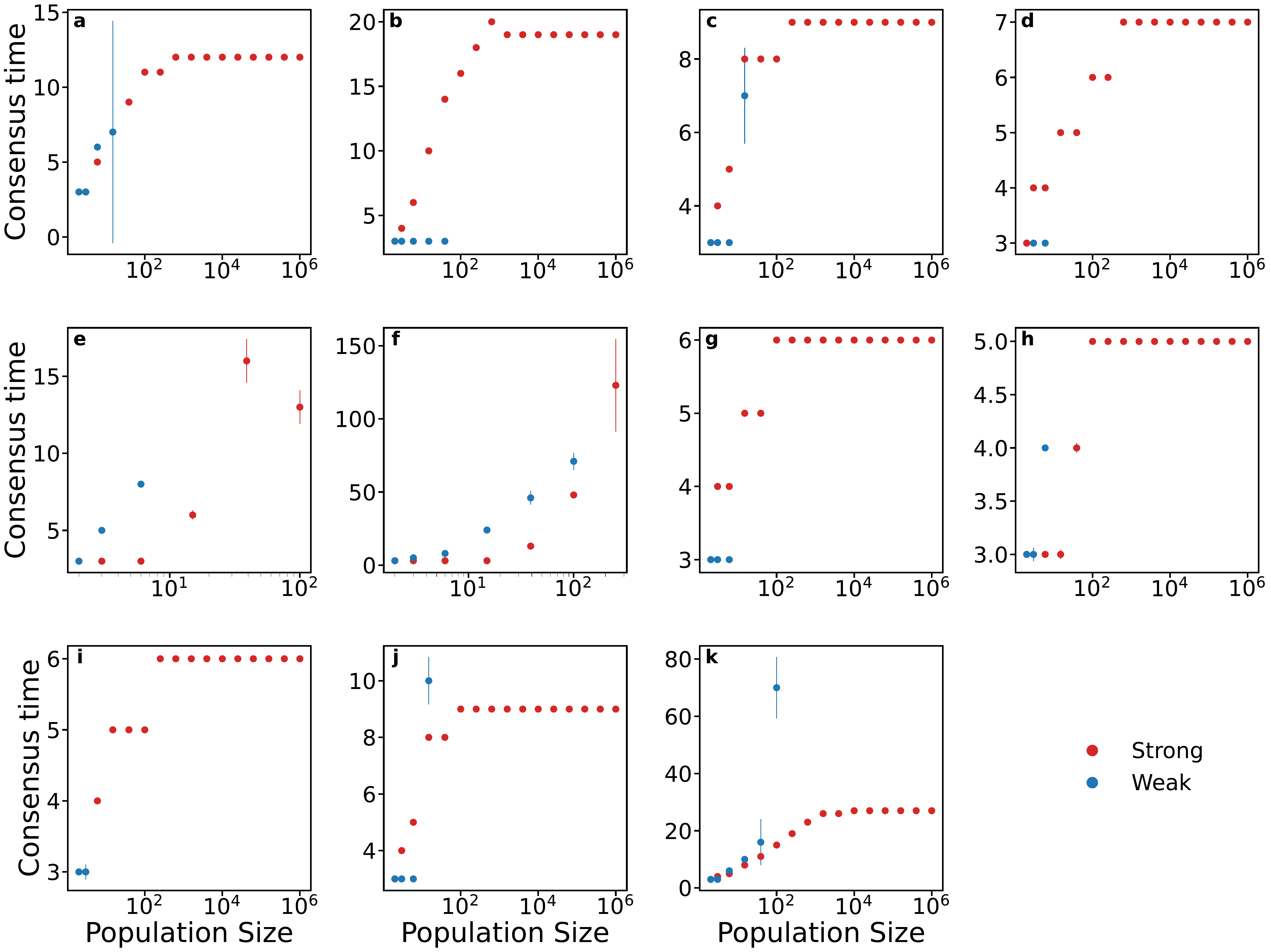}
    \caption{\textbf{Mode of the consensus time for Meta Llama 3.1 70B.} Each panel shows the mode of the consensus time for different word pairs as population size grows.}
    \label{fig: mode_time_llama}
\end{figure}

\begin{figure}[!h]
    \centering
    \includegraphics[width=\textwidth]{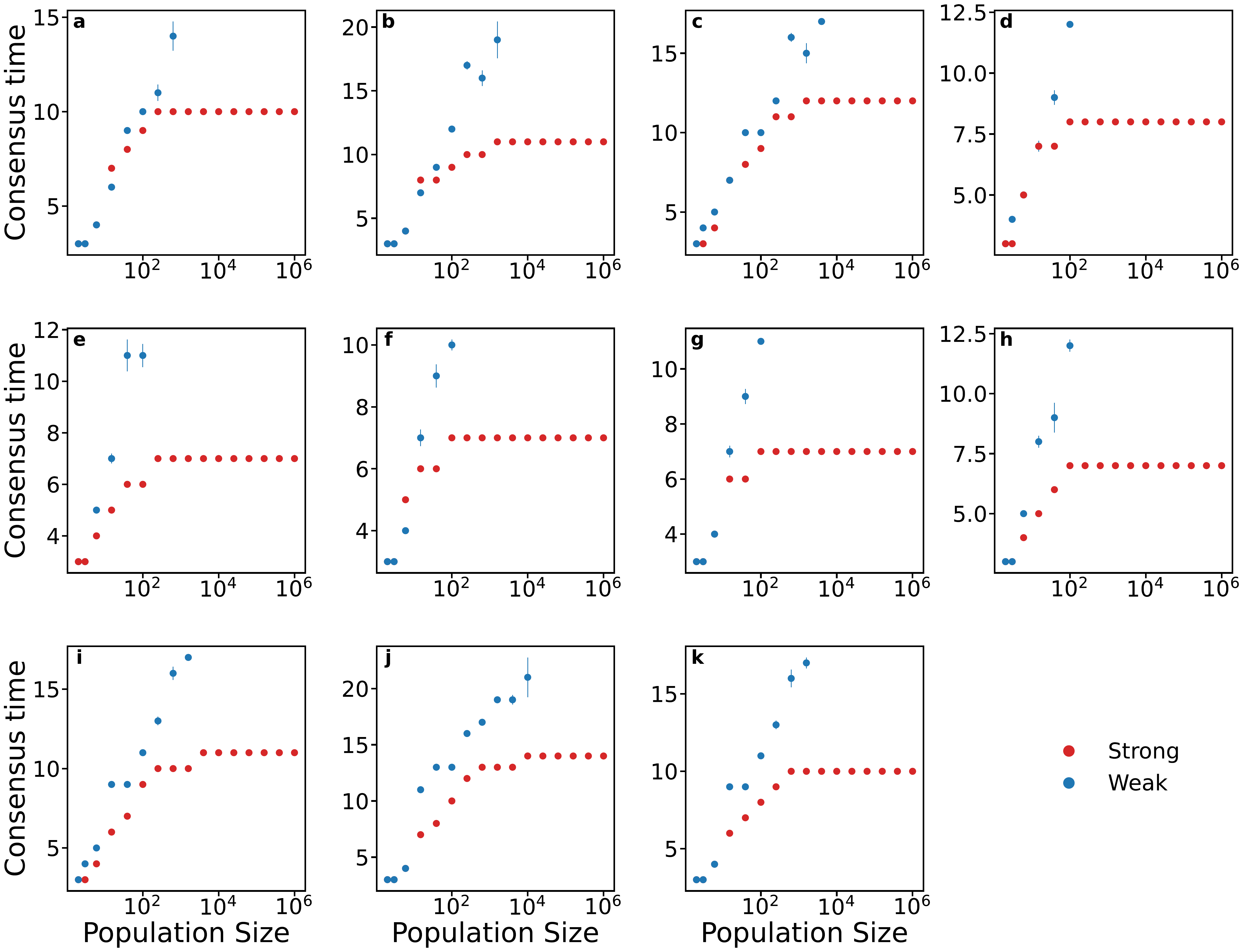}
    \caption{\textbf{Mode of the consensus time for  OpenAI GPT-4o.} Each panel shows the mode of the consensus time for different word pairs as population size grows.}
    \label{fig: mode_time_gpt}
\end{figure}

\begin{figure}[!h]
    \centering
    \includegraphics[width=\textwidth]{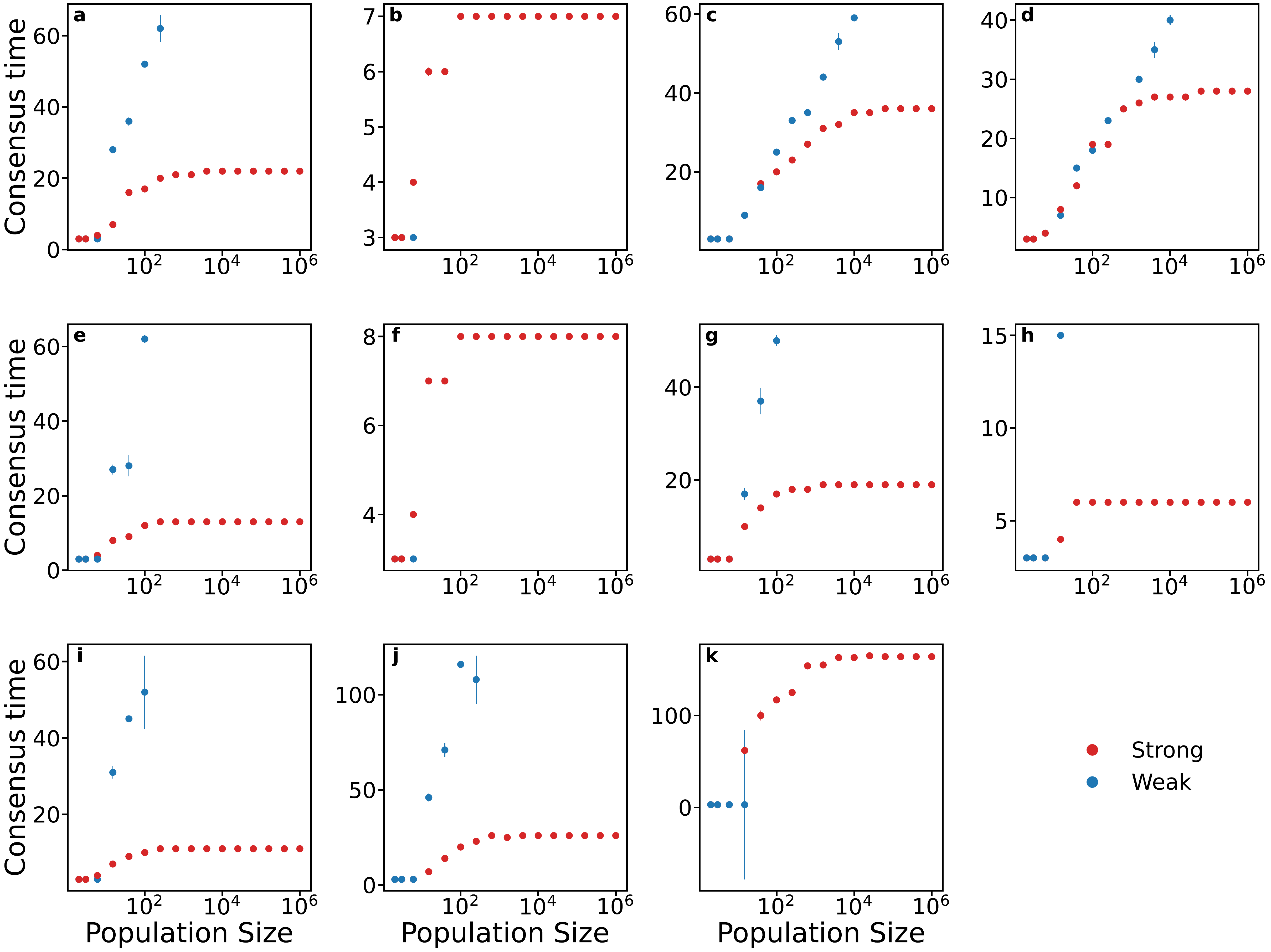}
    \caption{\textbf{Mode of the consensus time for  Qwen QwQ-32B.} Each panel shows the mode of the consensus time for different word pairs as population size grows.}
    \label{fig: mode_time_qwq}
\end{figure}

\begin{figure}[!h]
    \centering
    \includegraphics[width=\textwidth]{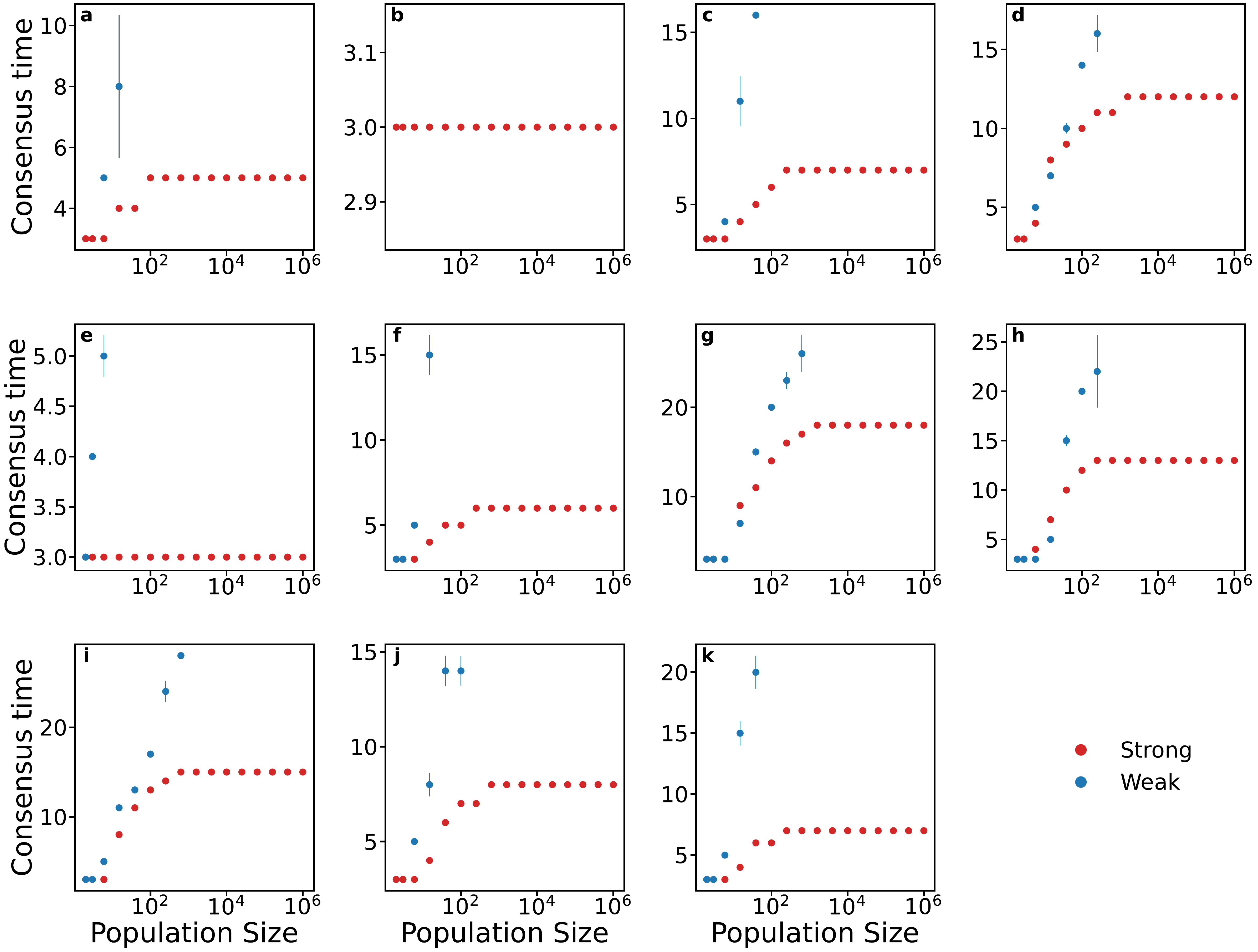}
    \caption{\textbf{Mode of the consensus time for  Microsoft Phi-4.} Each panel shows the mode of the consensus time for different word pairs as population size grows.}
    \label{fig: mode_time_phi}
\end{figure}

%% collective bias vs consensus time

\begin{figure}[!h]
    \centering
    \includegraphics[width=\textwidth]{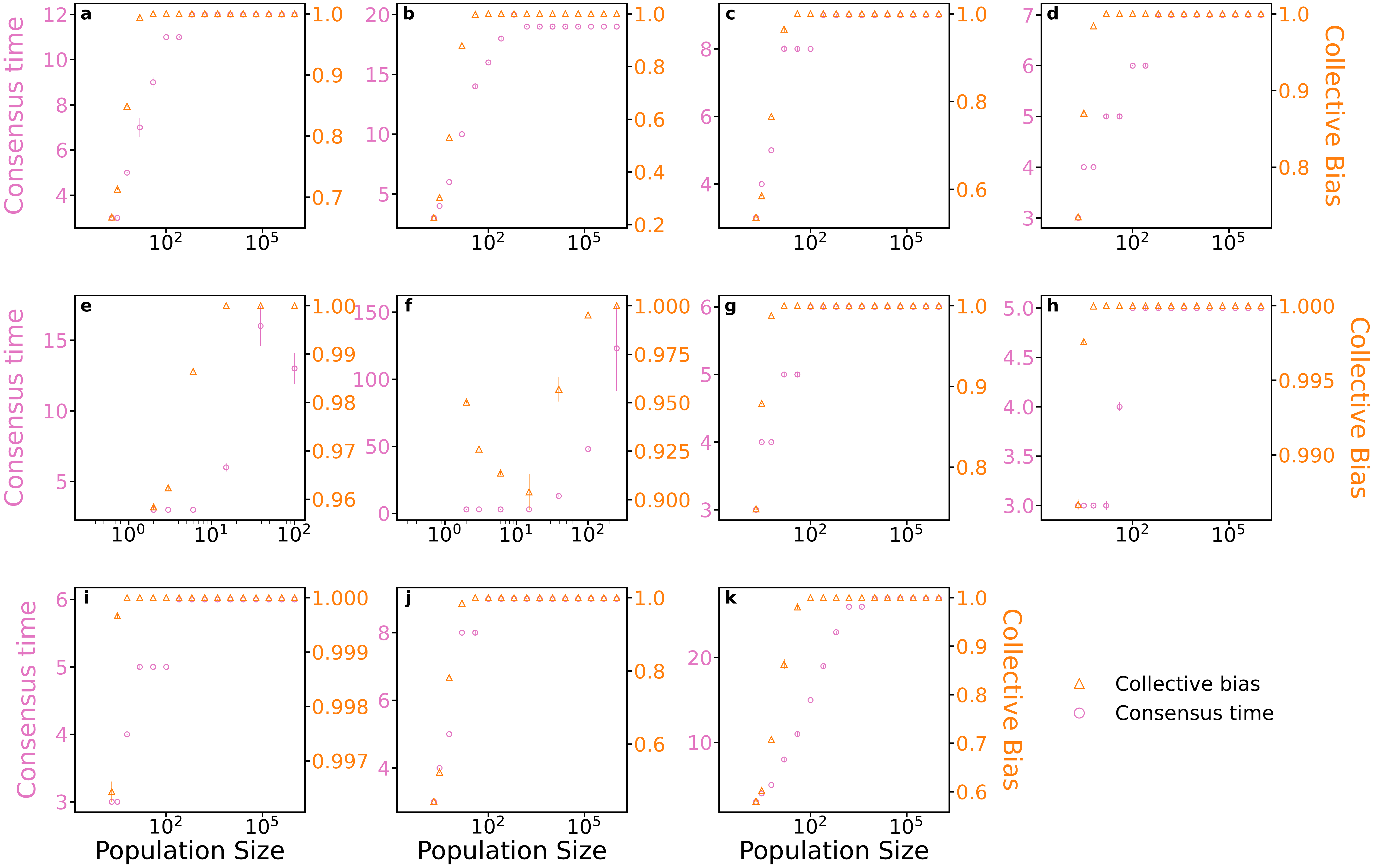}
    \caption{\textbf{Population size effect on collective bias and characteristic consensus time in Meta Llama 3.1 70B populations.} Each panel shows the change in collective bias (pink circles) and mode consensus time (orange triangles) as $N$ grows. Errors are standard error of the mean.}
    \label{fig: bias_time_llama}
\end{figure}

\begin{figure}[!h]
    \centering
    \includegraphics[width=\textwidth]{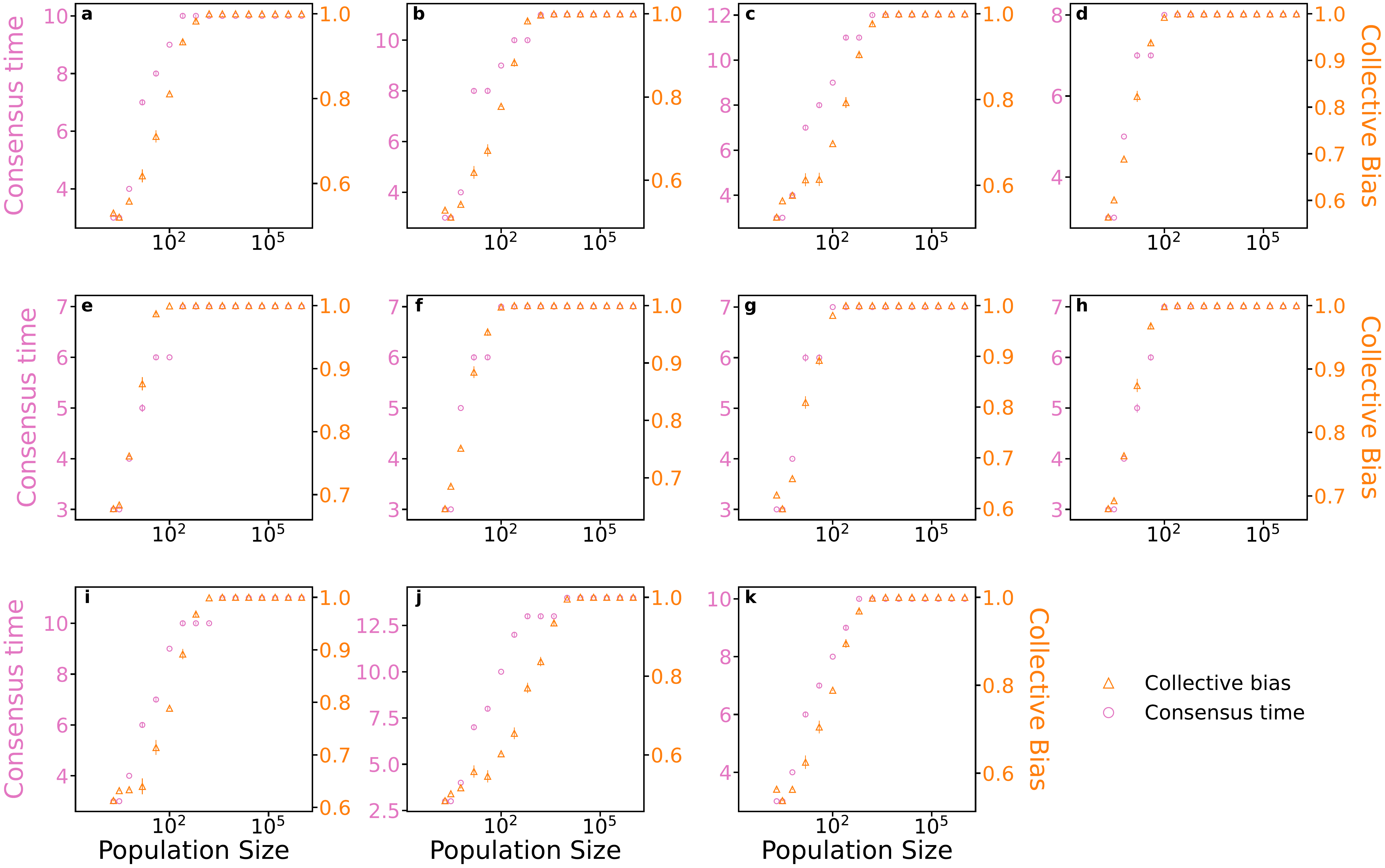}
    \caption{\textbf{Population size effect on collective bias and characteristic consensus time in OpenAI GPT-4o populations.} Each panel shows the change in collective bias (pink circles) and mode consensus time (orange triangles) as $N$ grows. Errors are standard error of the mean.}
    \label{fig: bias_time_gpt}
\end{figure}

\begin{figure}[!h]
    \centering
    \includegraphics[width=\textwidth]{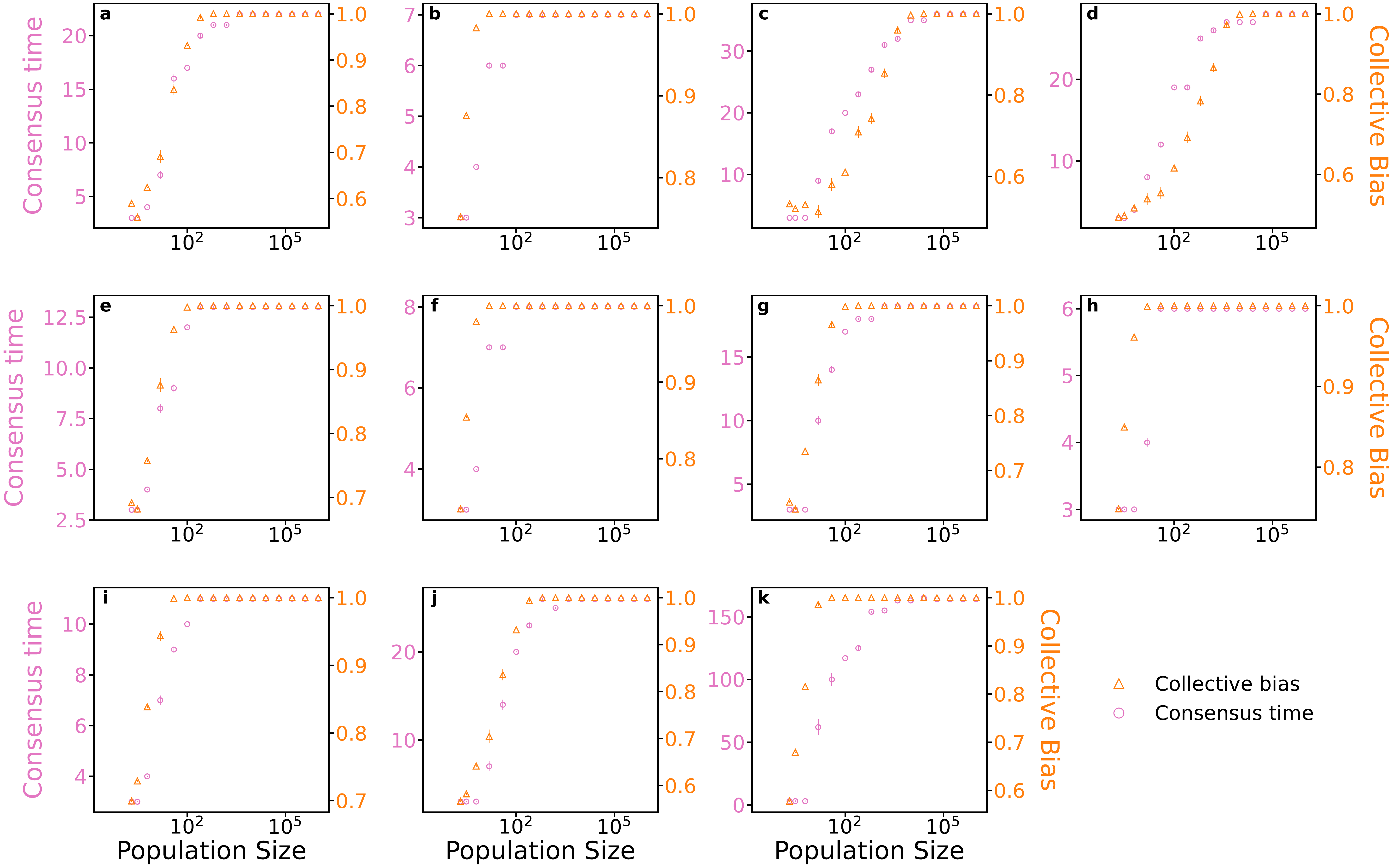}
    \caption{\textbf{Population size effect on collective bias and characteristic consensus time in Qwen QwQ-32B populations.} Each panel shows the change in collective bias (pink circles) and mode consensus time (orange triangles) as $N$ grows. Errors are standard error of the mean.}
    \label{fig: bias_time_qwq}
\end{figure}

\begin{figure}[!h]
    \centering
    \includegraphics[width=\textwidth]{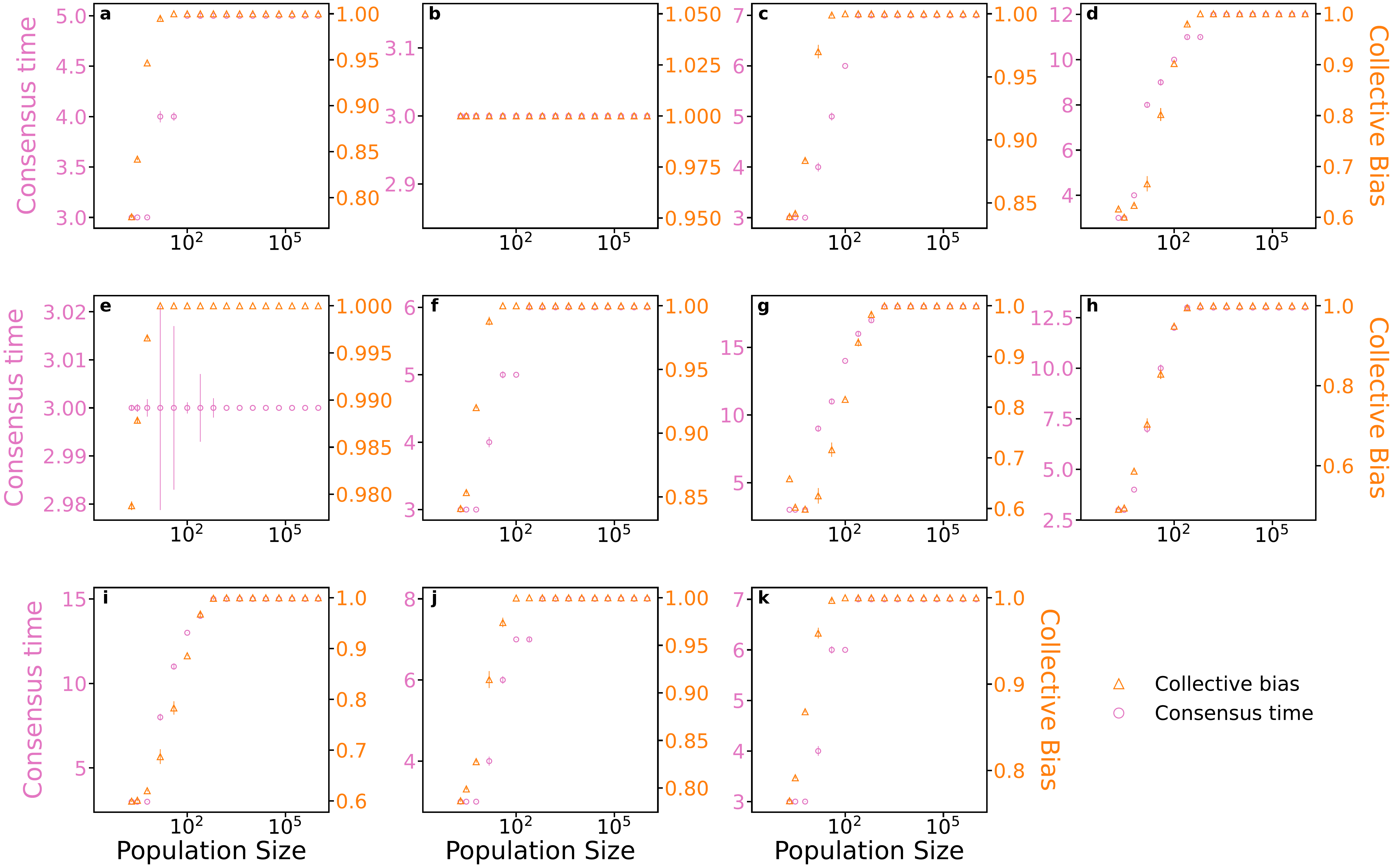}
    \caption{\textbf{Population size effect on collective bias and characteristic consensus time in Microsoft Phi-4 populations.} Each panel shows the change in collective bias (pink circles) and mode consensus time (orange triangles) as $N$ grows. Errors are standard error of the mean.}
    \label{fig: bias_time_phi}
\end{figure}

% consensus time PDFs

\begin{figure}[!h]
    \centering
    \includegraphics[width=\textwidth]{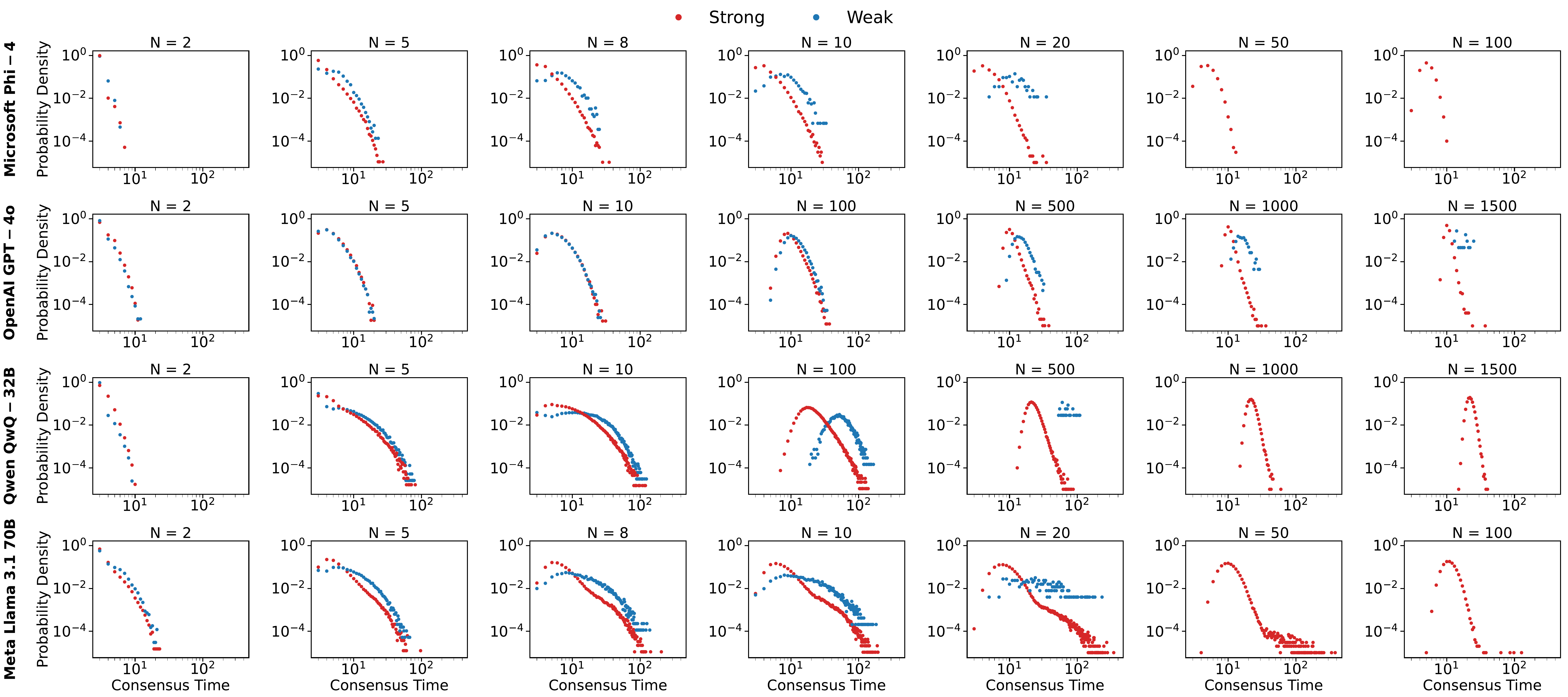}
    \caption{\textbf{PDF of the consensus time for  for the word pair \{her, his\}.} Each row corresponds to the transformation of the PDF as $N$ grows for a different LLM. Red and Blue circles correspond to trajectories that converged on the strong and weak word, respectively.}
    \label{fig: PDF_191}
\end{figure}

\begin{figure}[!h]
    \centering
    \includegraphics[width=\textwidth]{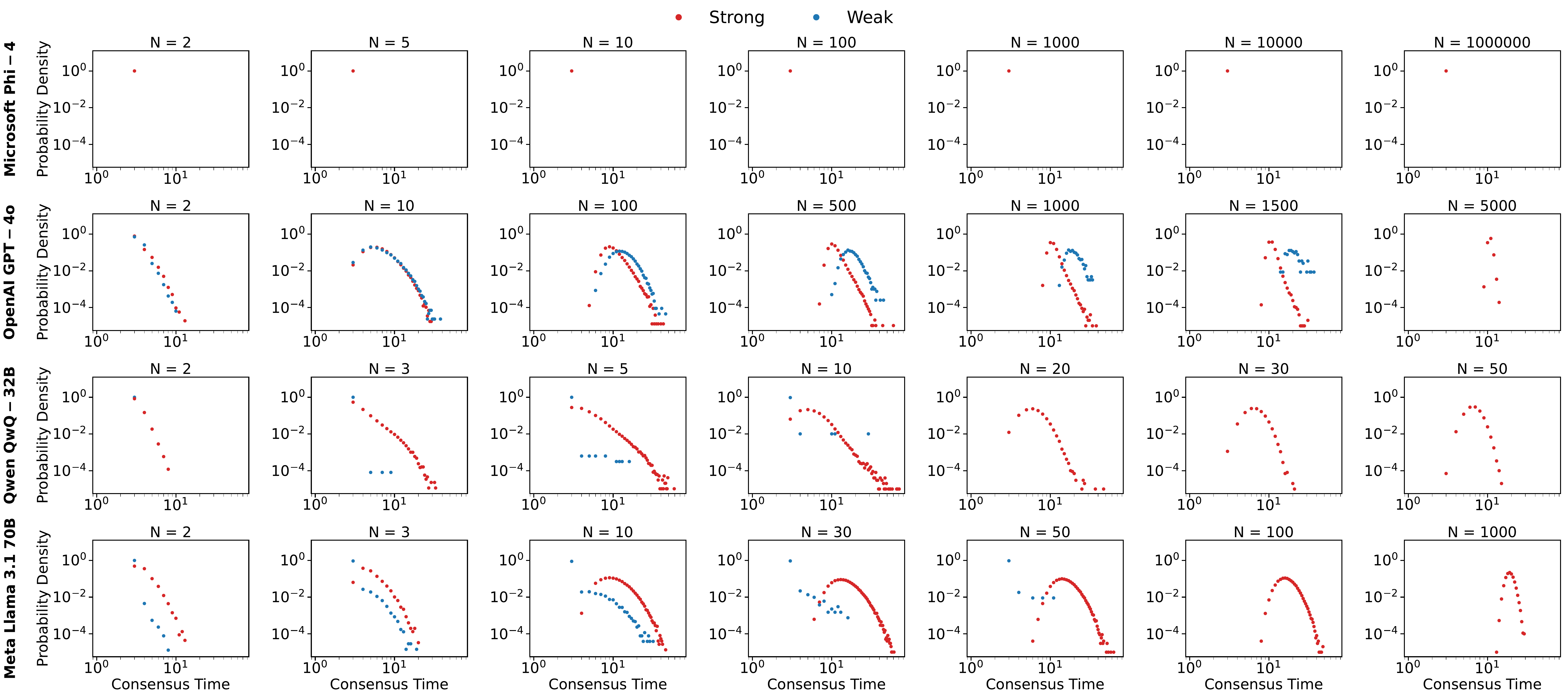}
    \caption{\textbf{PDF of the consensus time for  for the word pair \{straight, gay\}.} Each row corresponds to the transformation of the PDF as $N$ grows for a different LLM. Red and Blue circles correspond to trajectories that converged on the strong and weak word, respectively.}
    \label{fig: PDF_102}
\end{figure}

\begin{figure}[!h]
    \centering
    \includegraphics[width=\textwidth]{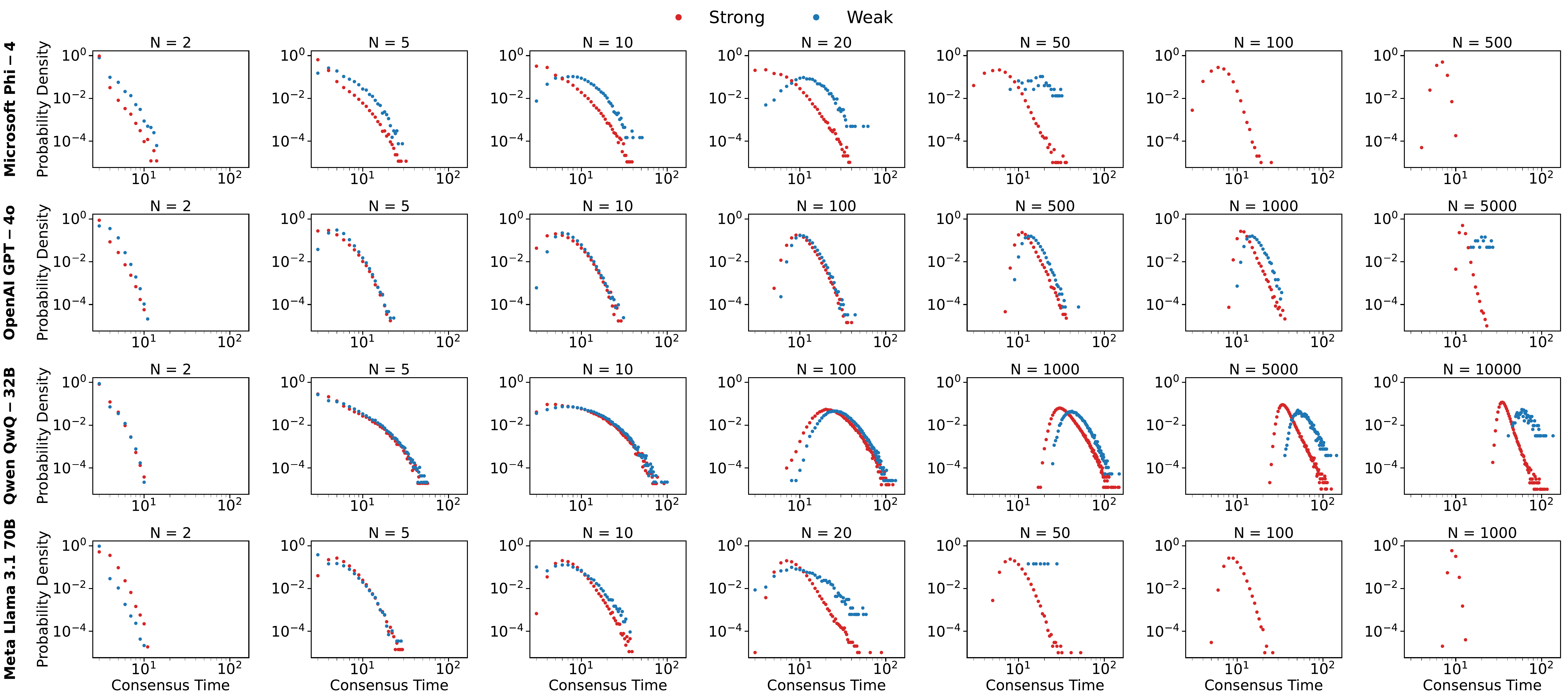}
    \caption{\textbf{PDF of the consensus time for  for the word pair \{he, she\}.} Each row corresponds to the transformation of the PDF as $N$ grows for a different LLM. Red and Blue circles correspond to trajectories that converged on the strong and weak word, respectively.}
    \label{fig: PDF_397}
\end{figure}

\begin{figure}[!h]
    \centering
    \includegraphics[width=\textwidth]{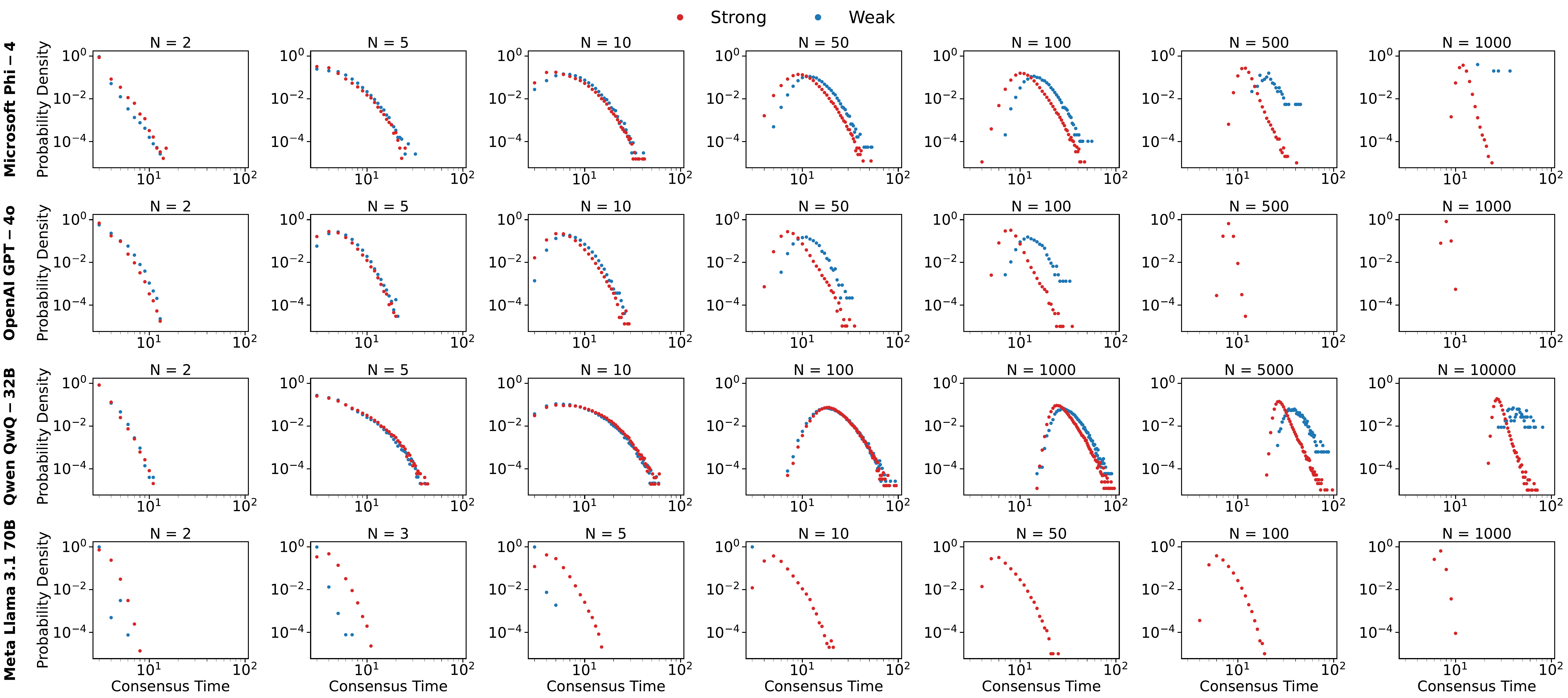}
    \caption{\textbf{PDF of the consensus time for  for the word pair \{Black, White\}.} Each row corresponds to the transformation of the PDF as $N$ grows for a different LLM. Red and Blue circles correspond to trajectories that converged on the strong and weak word, respectively.}
    \label{fig: PDF_344}
\end{figure}

\begin{figure}[!h]
    \centering
    \includegraphics[width=\textwidth]{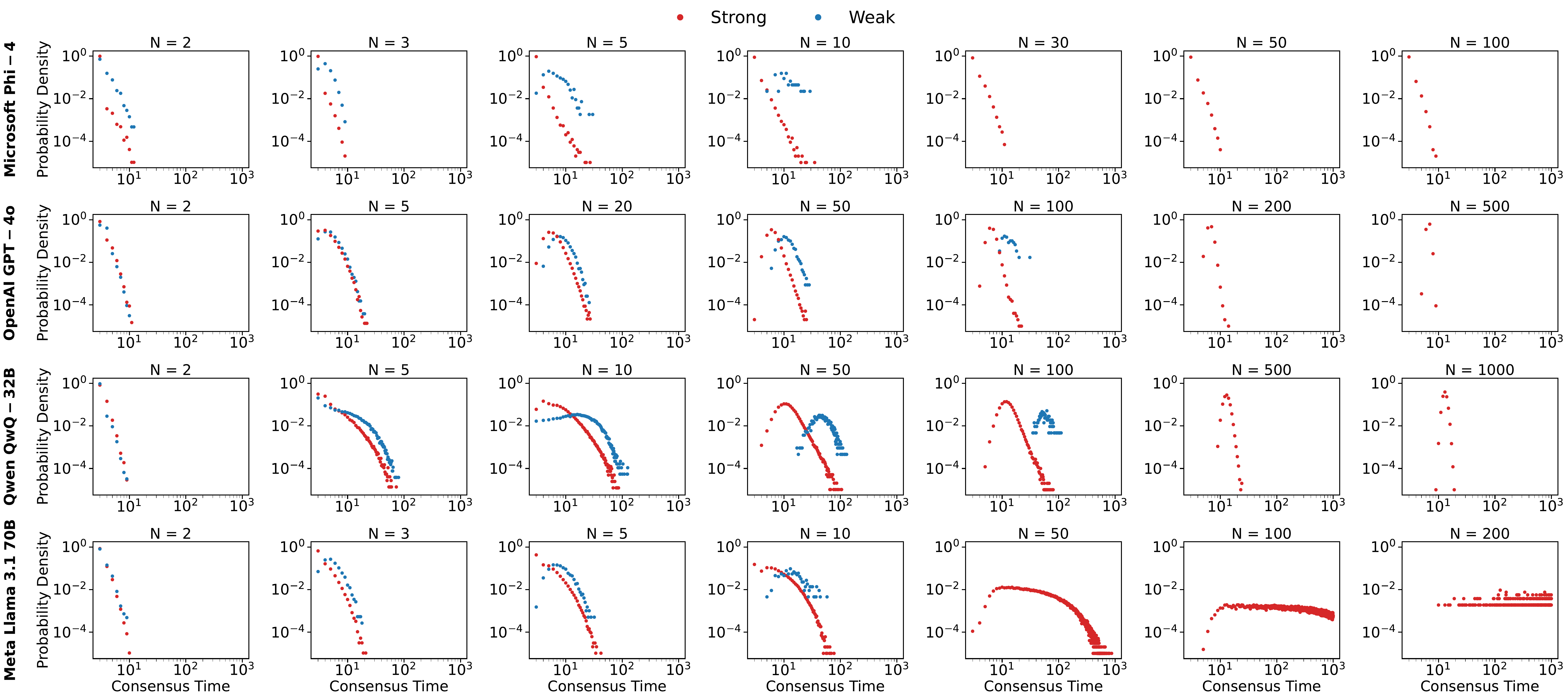}
    \caption{\textbf{PDF of the consensus time for  for the word pair \{less, more\}.} Each row corresponds to the transformation of the PDF as $N$ grows for a different LLM. Red and Blue circles correspond to trajectories that converged on the strong and weak word, respectively.}
    \label{fig: PDF_1161}
\end{figure}

\begin{figure}[!h]
    \centering
    \includegraphics[width=\textwidth]{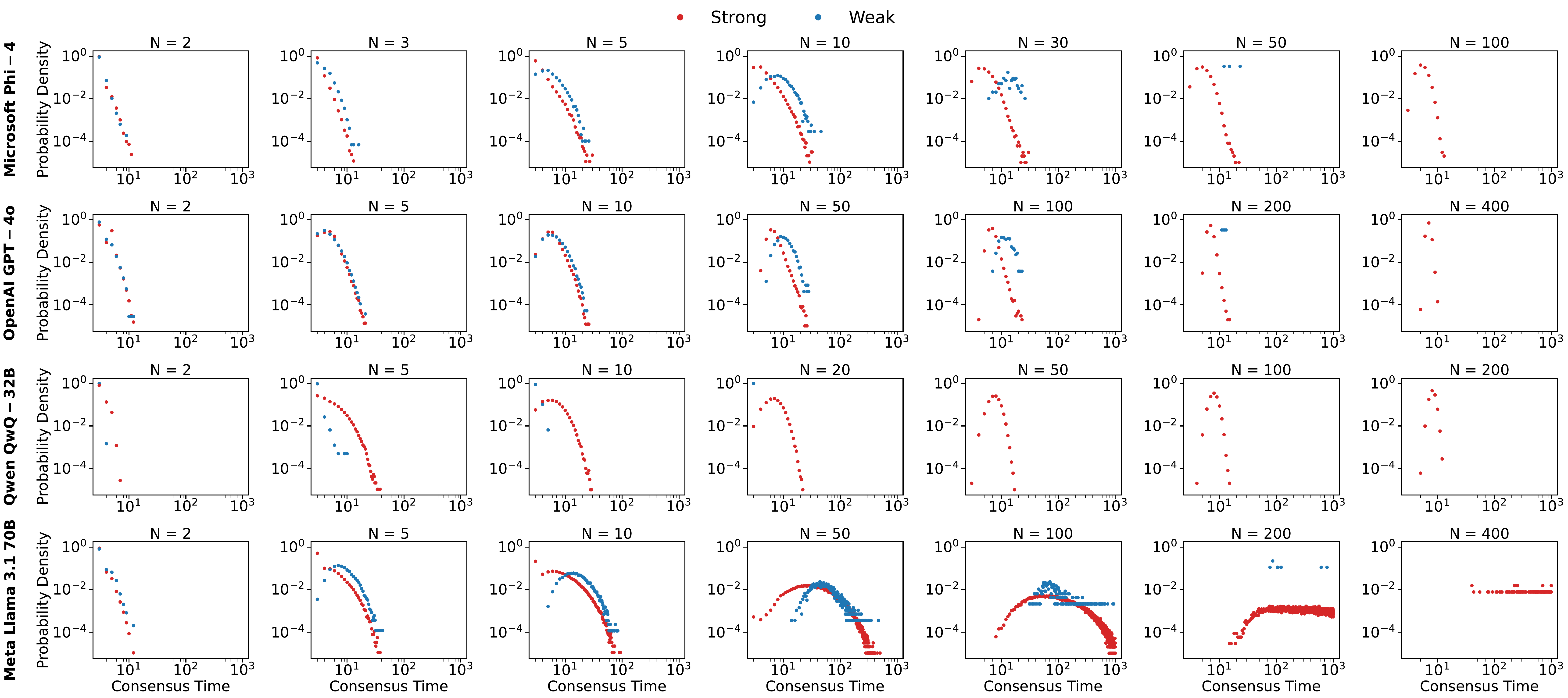}
    \caption{\textbf{PDF of the consensus time for  for the word pair \{old, young\}.} Each row corresponds to the transformation of the PDF as $N$ grows for a different LLM. Red and Blue circles correspond to trajectories that converged on the strong and weak word, respectively.}
    \label{fig: PDF_219}
\end{figure}

\begin{figure}[!h]
    \centering
    \includegraphics[width=\textwidth]{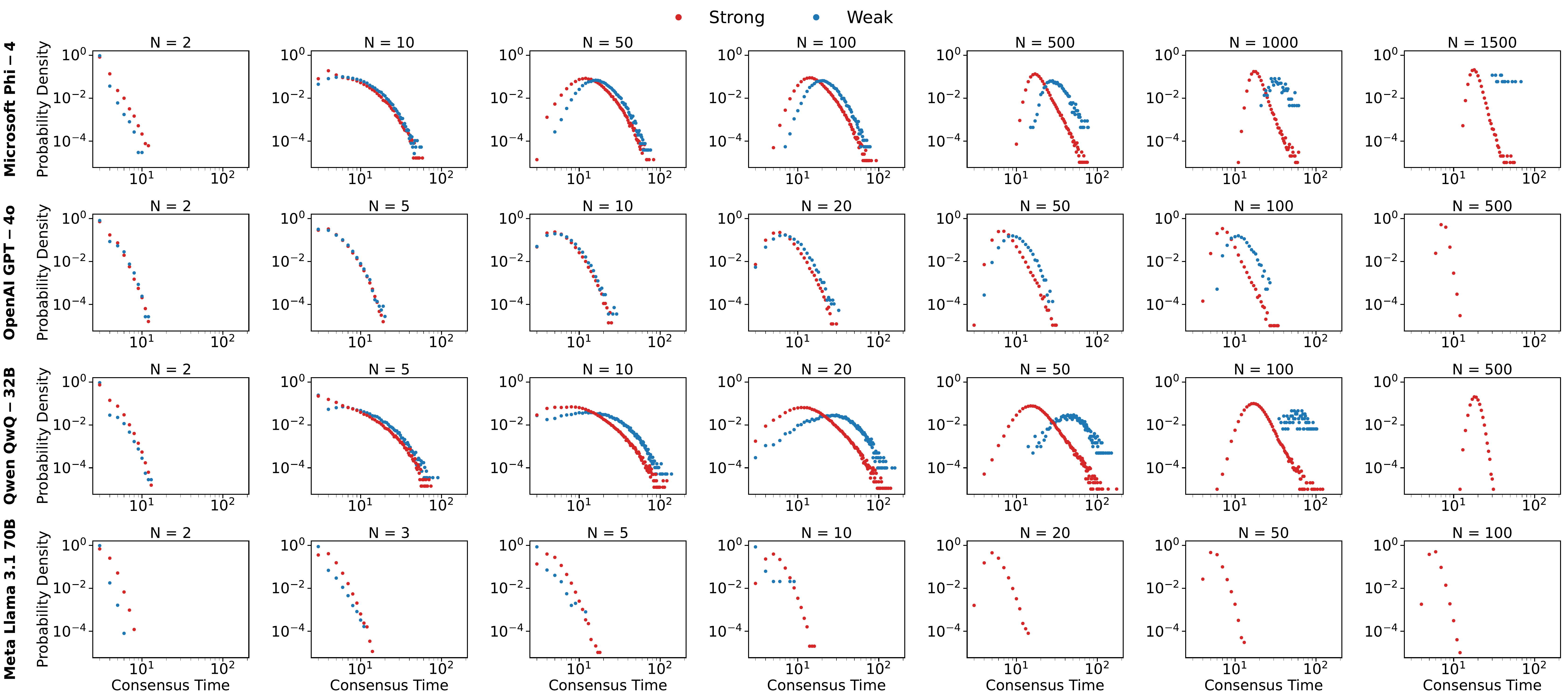}
    \caption{\textbf{PDF of the consensus time for  for the word pair \{White, African\}.} Each row corresponds to the transformation of the PDF as $N$ grows for a different LLM. Red and Blue circles correspond to trajectories that converged on the strong and weak word, respectively.}
    \label{fig: PDF_92}
\end{figure}

\begin{figure}[!h]
    \centering
    \includegraphics[width=\textwidth]{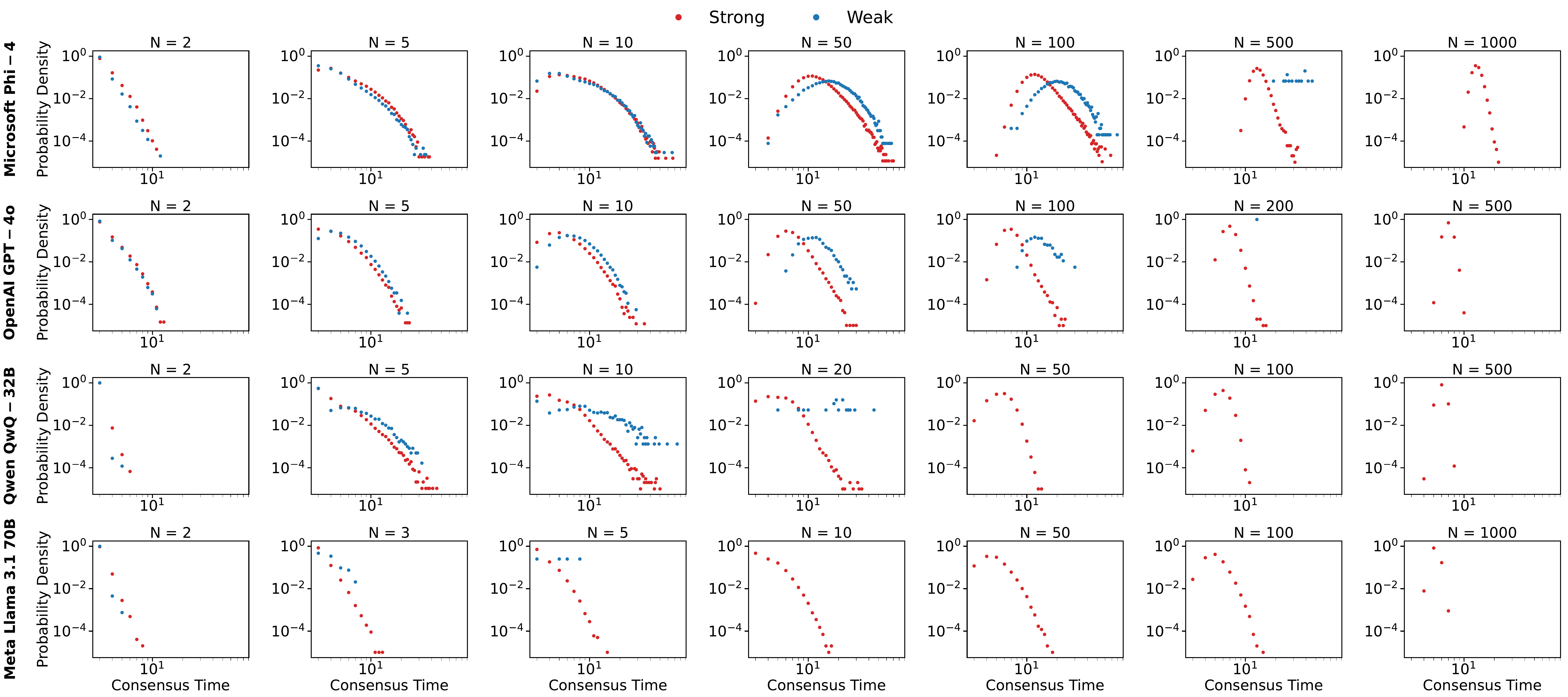}
    \caption{\textbf{PDF of the consensus time for  for the word pair \{American, Mexican\}.} Each row corresponds to the transformation of the PDF as $N$ grows for a different LLM. Red and Blue circles correspond to trajectories that converged on the strong and weak word, respectively.}
    \label{fig: PDF_349}
\end{figure}

\begin{figure}[!h]
    \centering
    \includegraphics[width=\textwidth]{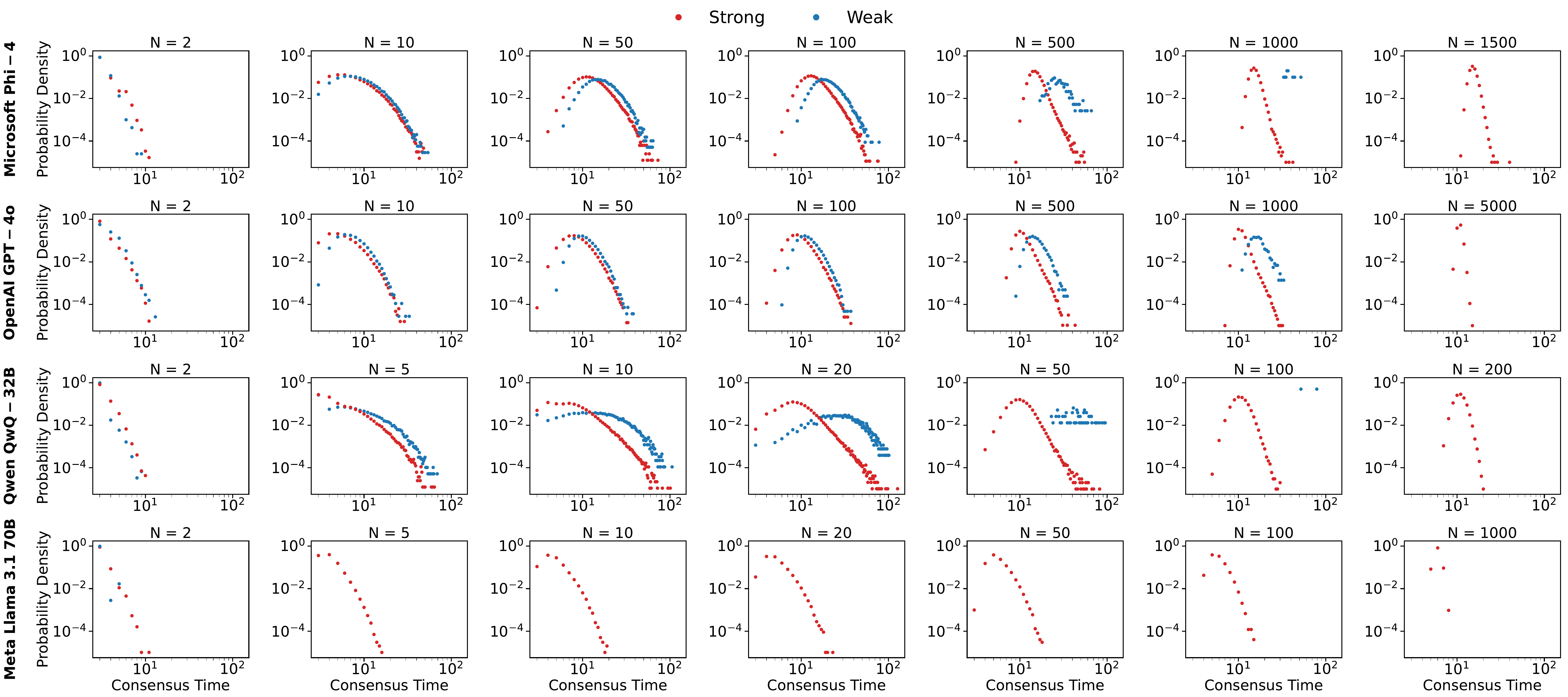}
    \caption{\textbf{PDF of the consensus time for  for the word pair \{short, tall\}.} Each row corresponds to the transformation of the PDF as $N$ grows for a different LLM. Red and Blue circles correspond to trajectories that converged on the strong and weak word, respectively.}
    \label{fig: PDF_626}
\end{figure}

\begin{figure}[!h]
    \centering
    \includegraphics[width=\textwidth]{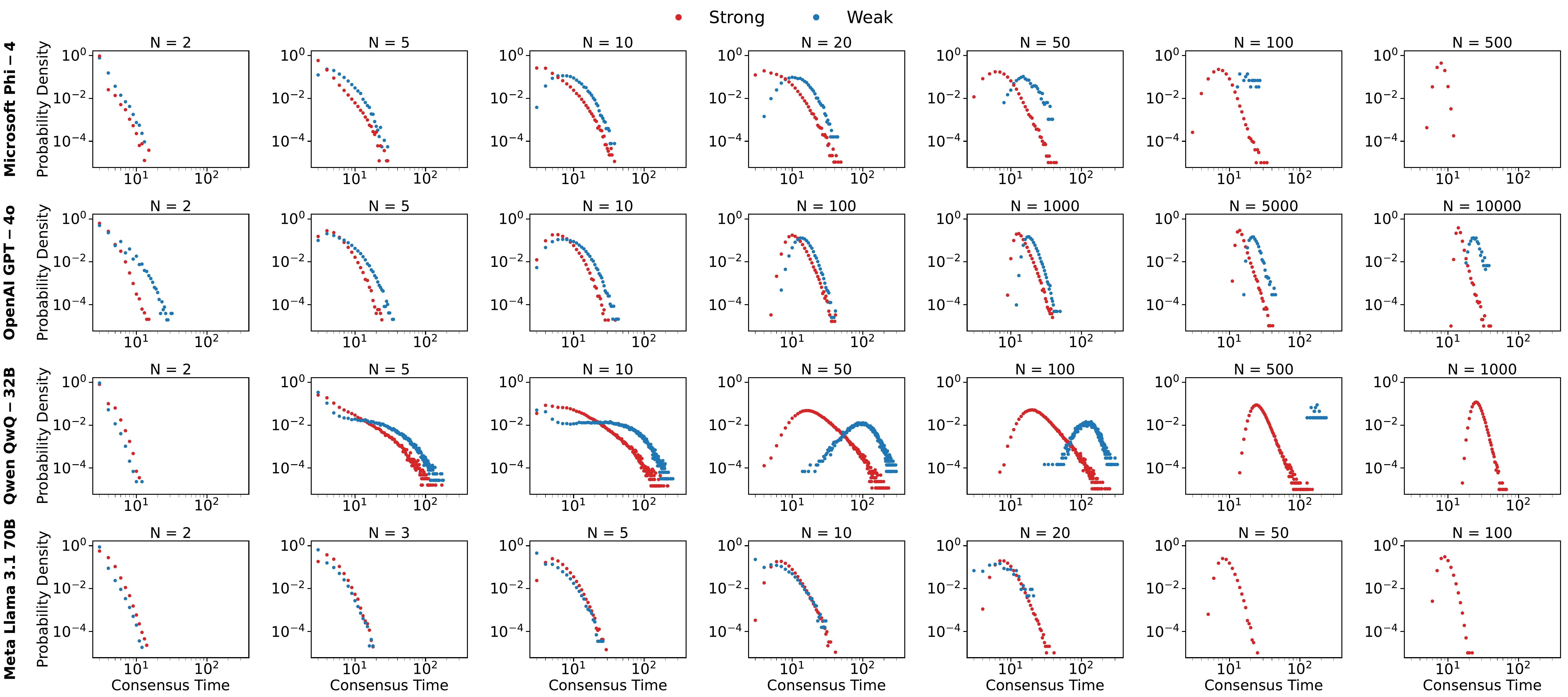}
    \caption{\textbf{PDF of the consensus time for  for the word pair \{man, woman\}.} Each row corresponds to the transformation of the PDF as $N$ grows for a different LLM. Red and Blue circles correspond to trajectories that converged on the strong and weak word, respectively.}
    \label{fig: PDF_505}
\end{figure}

\begin{figure}[!h]
    \centering
    \includegraphics[width=\textwidth]{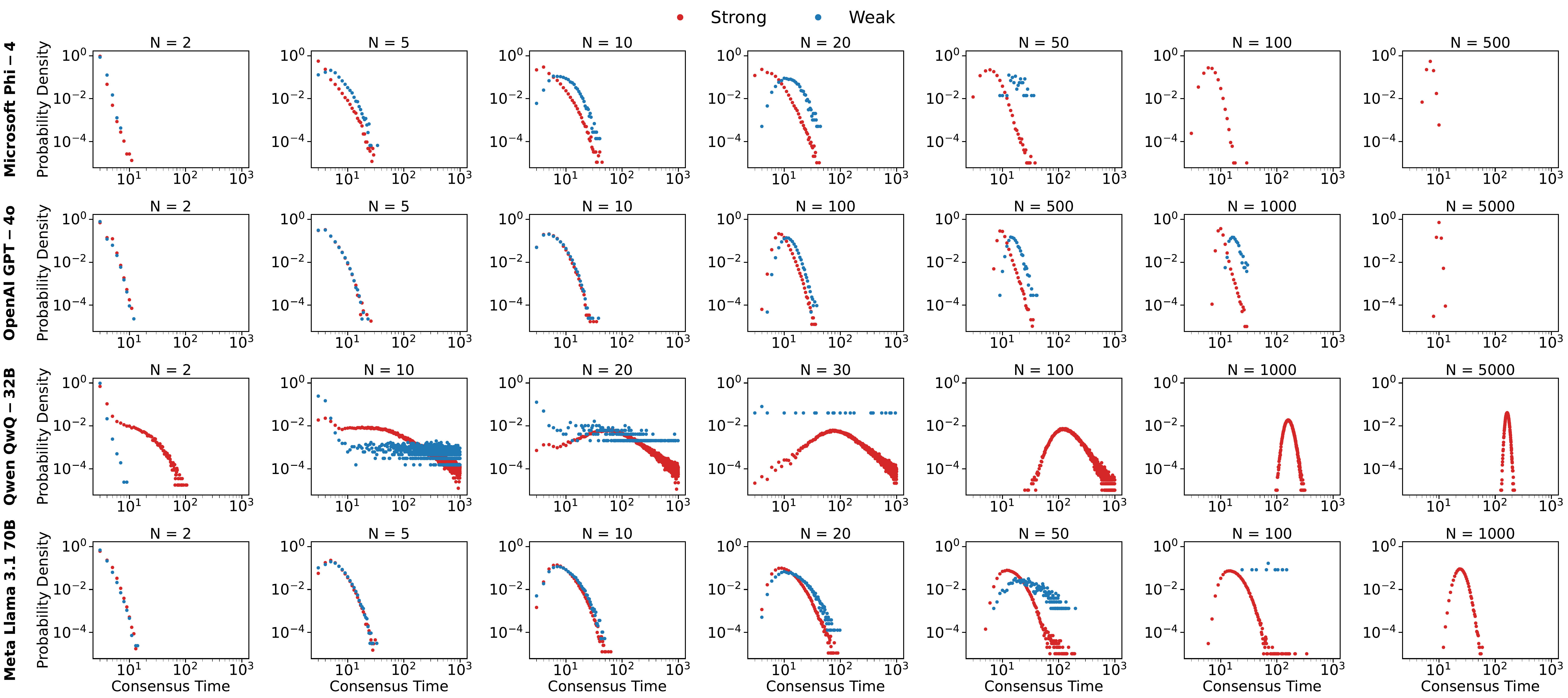}
    \caption{\textbf{PDF of the consensus time for  for the word pair \{husband, wife\}.} Each row corresponds to the transformation of the PDF as $N$ grows for a different LLM. Red and Blue circles correspond to trajectories that converged on the strong and weak word, respectively.}
    \label{fig: PDF_410}
\end{figure}

\begin{figure}[!h]
    \centering
    \includegraphics[width=\textwidth]{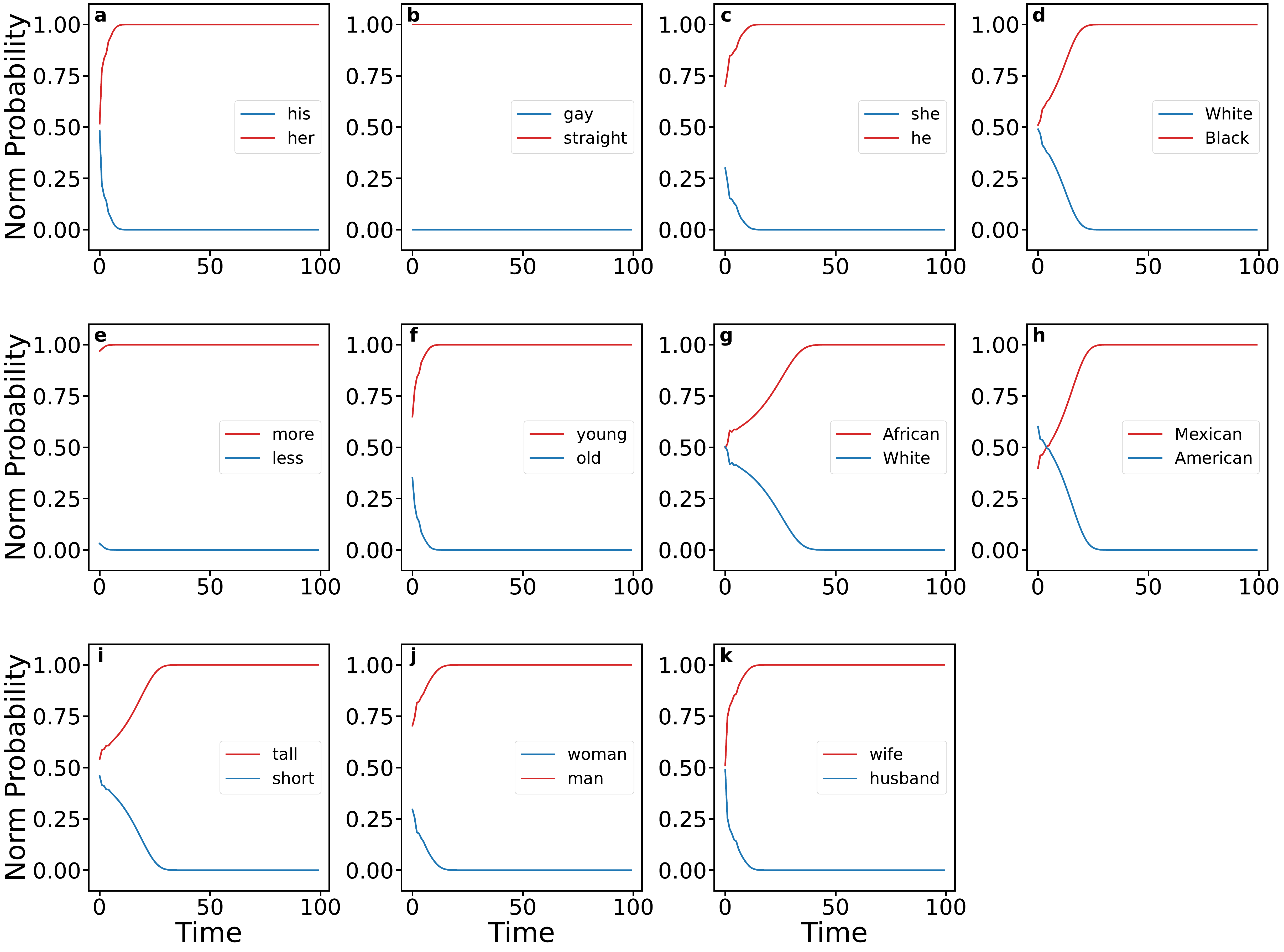}
    \caption{\textbf{Mean-field dynamics for Microsoft Phi-4 populations.} Each panel shows the temporal evolution of word probabilities under mean-field dynamics for the corresponding word pair.}
    \label{fig: mean_field_phi}
\end{figure}

\begin{figure}[!h]
    \centering
    \includegraphics[width=\textwidth]{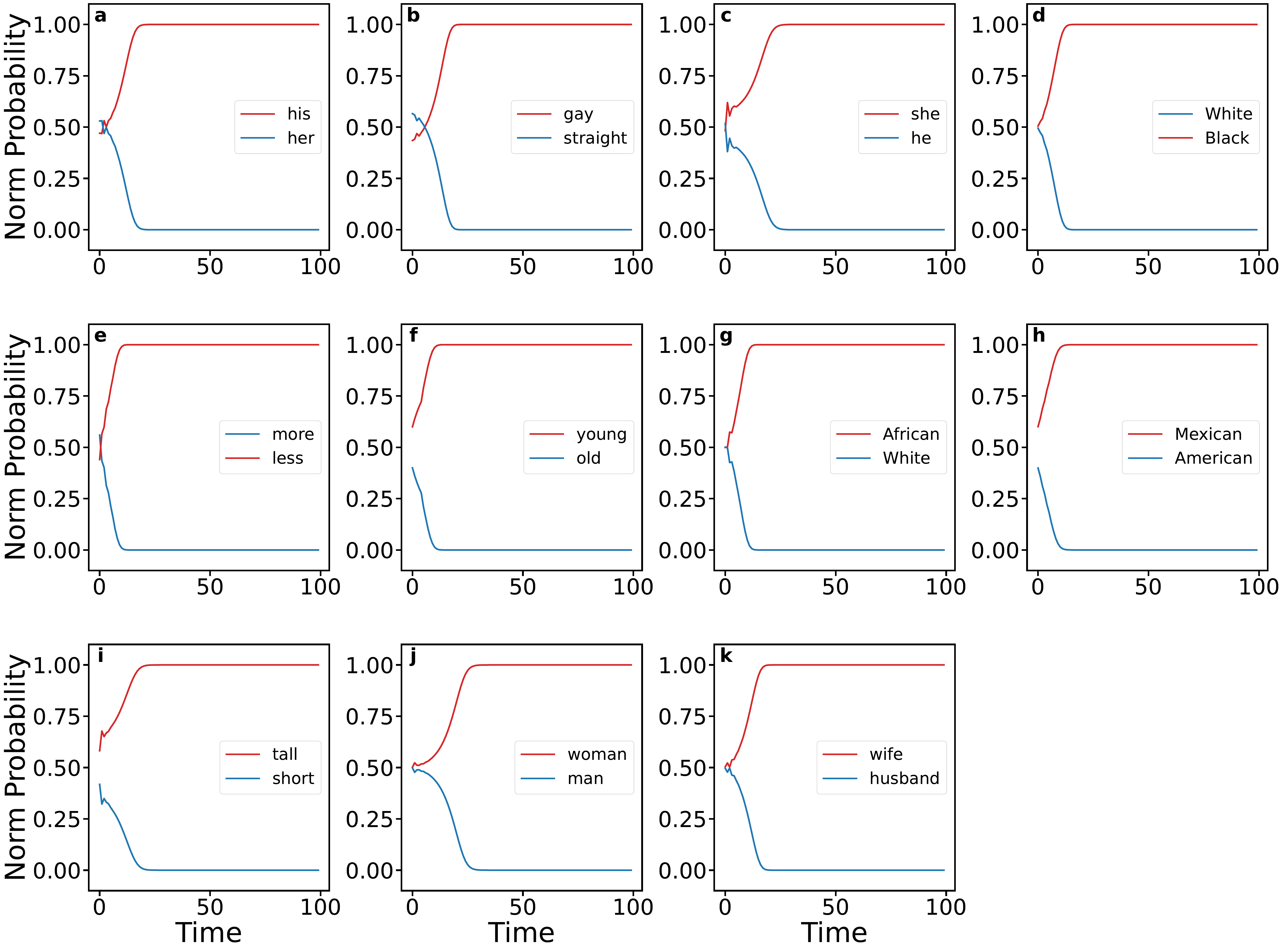}
    \caption{\textbf{Mean-field dynamics for OpenAI gpt-4o populations.} Each panel shows the temporal evolution of word probabilities under mean-field dynamics for the corresponding word pair.}
    \label{fig: mean_field_gpt}
\end{figure}
\begin{figure}[!h]
    \centering
    \includegraphics[width=\textwidth]{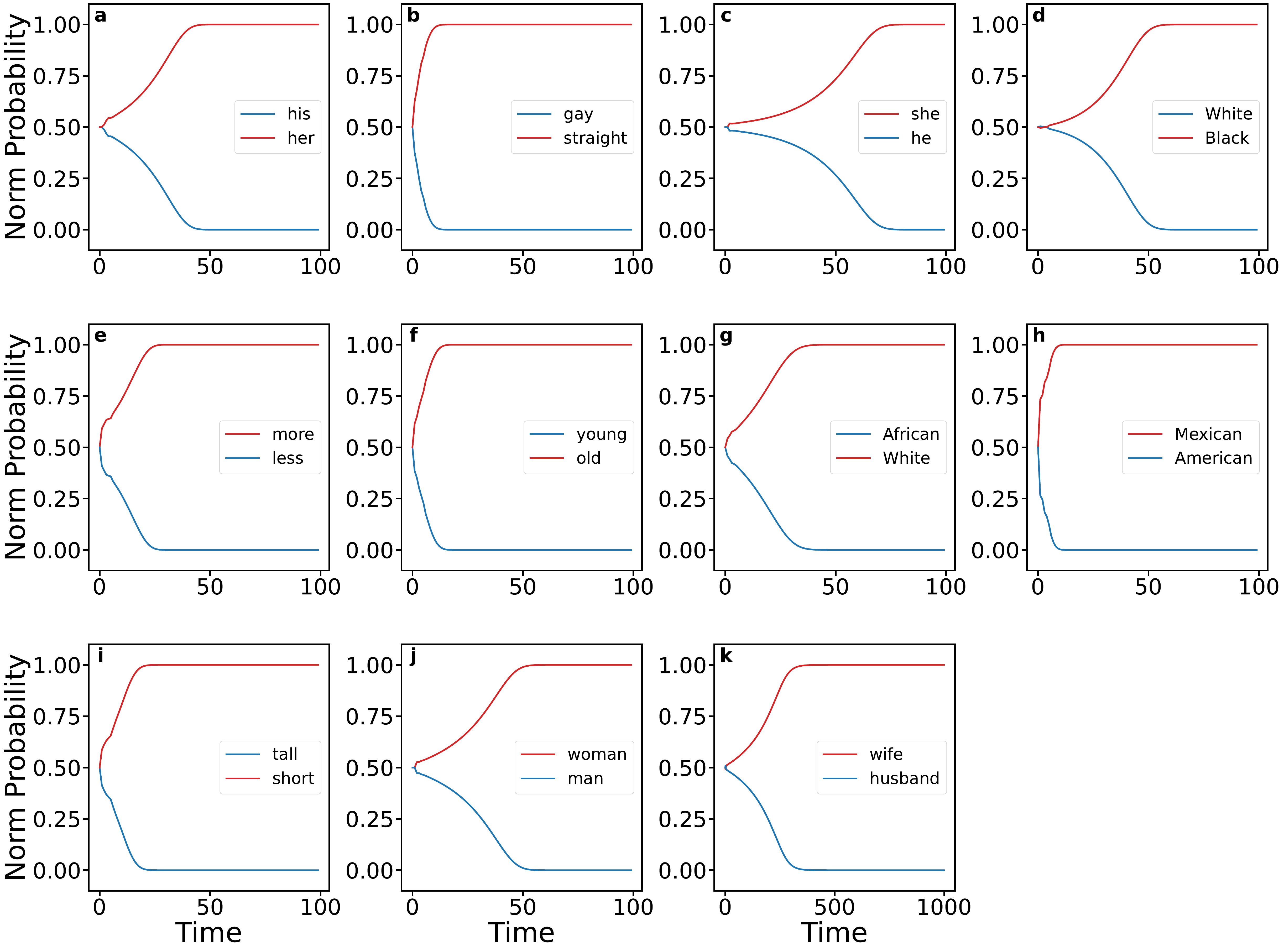}
    \caption{\textbf{Mean-field dynamics for Qwen QwQ-32B populations.} Each panel shows the temporal evolution of word probabilities under mean-field dynamics for the corresponding word pair.}
    \label{fig: mean_field_qwq}
\end{figure}

\begin{figure}[!h]
    \centering
    \includegraphics[width=\textwidth]{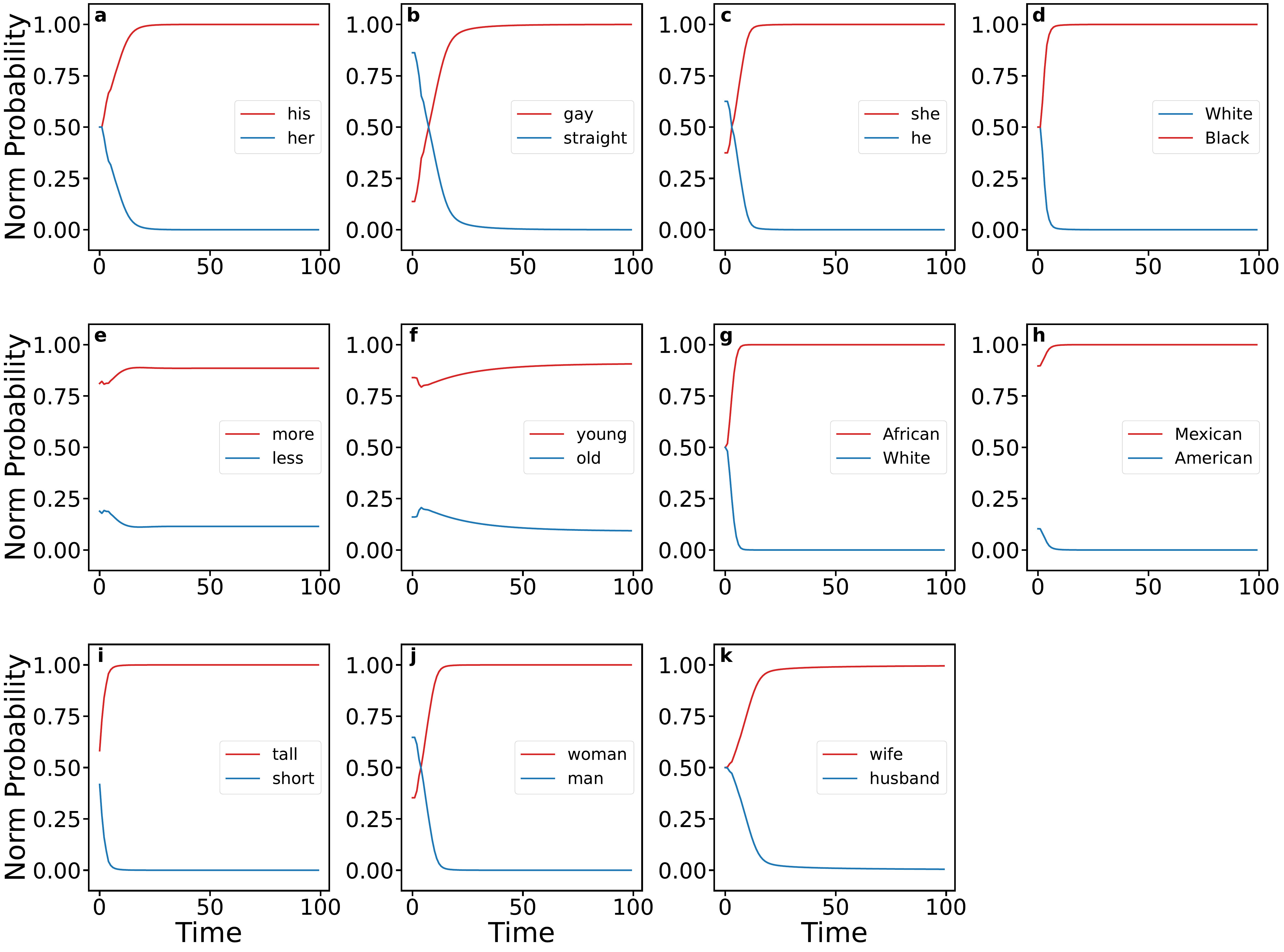}
    \caption{\textbf{Mean-field dynamics for Meta Llama 3.1 70B populations.} Each panel shows the temporal evolution of word probabilities under mean-field dynamics for the corresponding word pair.}
    \label{fig: mean_field_llama}
\end{figure}

\begin{table}%[!h]
\centering
\resizebox{\textwidth}{!}{
\begin{tabular}{|l|cc|cc|cc|cc|}
\hline
\textbf{Word Pair} & \multicolumn{2}{c|}{\textbf{Qwen QwQ-32B}} & \multicolumn{2}{c|}{\textbf{OpenAI GPT-4o}} & \multicolumn{2}{c|}{\textbf{Meta Llama 3.1 70B}} & \multicolumn{2}{c|}{\textbf{Microsoft Phi-4}} \\
\cline{2-9}
& $\lambda_a$ & $\lambda_b$ & $\lambda_a$ & $\lambda_b$ & $\lambda_a$ & $\lambda_b$ & $\lambda_a$ & $\lambda_b$ \\
\hline
her$_a$, his$_b$ & \textbf{-0.543} & -0.101 & -0.766 & \textbf{-0.615} & -0.234 & \textbf{-0.202} & \textbf{-0.933} & -0.553 \\
\hline
straight$_a$, gay$_b$ & \textbf{-0.614} & 0.0180 & -0.279 & \textbf{-0.818} & 0.255 & \textbf{-0.060} & \textbf{-0.845} & -0.987 \\
\hline
he$_a$, she$_b$ & -0.300 & \textbf{-0.399} & -0.830 & \textbf{-0.574} & 0.057 & \textbf{-0.183} & \textbf{-0.904} &  -0.966\\
\hline
Black$_a$, White$_b$ & \textbf{-0.419} & -0.577 & \textbf{-0.848} & -0.345 & \textbf{-0.195} & 0.723 & \textbf{-0.662} & -0.550 \\
\hline
less$_a$, more$_b$ & -0.070 & \textbf{-0.850} & \textbf{-0.880} & -0.664 & 0.207 & \textbf{0.169} & -0.952 & \textbf{-0.805} \\
\hline
old$_a$, young$_b$ & \textbf{-8.50} & 0.265 & -0.738 & \textbf{-0.940} & -7.39e-16 & \textbf{0.0481} & -0.591 & \textbf{-0.952} \\
\hline
White$_a$, African$_b$ & \textbf{-0.346} & -0.089 & -0.616 & \textbf{-0.918} & 0.433 & \textbf{-0.453} & -0.423 & \textbf{-0.507} \\
\hline
American$_a$, Mexican$_b$ & -0.0702 & \textbf{-0.990} & -0.384 & \textbf{-0.399} & 0.297 & \textbf{-0.294} & -0.252 & \textbf{-0.640} \\
\hline
short$_a$, tall$_b$ & \textbf{-0.558} & -0.070 & -0.904 & \textbf{-0.530} & -0.0706 & \textbf{-0.276} & -0.879 & \textbf{-0.713} \\
\hline
man$_a$, woman$_b$ & -0.028 & \textbf{-0.447} & -0.135 & \textbf{-0.498} & 0.255 & \textbf{-0.252} & \textbf{-0.650} & -0.946 \\
\hline
husband$_a$, wife$_b$ & -4.58e-16 & \textbf{-0.041} & -0.402 & \textbf{-0.892} & 0.092 & \textbf{-2.39e-15} & -0.882 & \textbf{-0.921} \\
\hline
\end{tabular}%
}
\caption{\textbf{First eigenvalues (up to 3 s.f.) for the trivial candidate solutions of the mean-field dynamics across four LLMs.} Bold values indicate the strong word in each pair.}
\label{tab:combined_eigenvalues}
\end{table}

\begin{table}%[!h]
\begin{center}
\begin{tabular}{c p{2cm} p{8cm}}
 & \textbf{Name} & \multicolumn{1}{c}{\textbf{Question}} \\
\hline
\rotatebox[origin=c]{90}{\parbox[c]{1.5cm}{\centering Rules}} 
& \texttt{min\_max} & What is the lowest/highest payoff player A can get in a single round? \\ 
\cline{2-3}
& \texttt{actions} & Which actions is player A allowed to play? \\ 
\cline{2-3}
& \texttt{payoff} & Which is player X's payoff in a single round if $X$ plays $p$ and $Y$ plays $q$? \\
\hline
\rotatebox[origin=c]{90}{\parbox[c]{1.5cm}{\centering Time}} 
& \texttt{round} & Which is the current round of the game? \\
\cline{2-3}
& \texttt{action$_i$} & Which action did player $X$ play in round $i$? \\
\cline{2-3}
& \texttt{points$_i$} & How many points did player $X$ collect in round $i$? \\
\hline
\rotatebox[origin=c]{90}{\parbox[c]{1.5cm}{\centering State}} 
& \texttt{\#actions} & How many times did player $X$ choose $p$? \\
\cline{2-3}
& \texttt{\#points} & What is player $X$'s current total payoff? \\
\end{tabular}
\end{center}
\caption{\textbf{Meta-prompting questions.} Templates of prompt comprehension questions used in meta-prompting to verify the LLM's comprehension of the prompt.}
\label{tab:meta-prompting}
\end{table}

\end{document}